\documentstyle[11pt]{article}
\newcommand{\blankline}{\vskip .3cm}
\newcommand{\f}{\begin{equation}}
\newcommand{\ff}{\end{equation}}

\makeatletter                                                           

\def\section{\@startsection {section}{1}{\z@}{-1.5ex plus -.5ex         
minus -.2ex}{1ex plus .2ex}{\large\bf}}                                 

\def\@thmcountersep{}                                                   

\long\def\@makecaption#1#2{\vskip 10pt \setbox\@tempboxa\hbox{#1. #2}   
   \ifdim \wd\@tempboxa >\hsize   
       #1. #2\par                 
     \else                        
       \hbox to\hsize{\hfil\box\@tempboxa\hfil}                         
   \fi}                                                                 

\def\ps@headings{                                                       
 \def\@oddhead{\footnotesize\rm\hfill\runninghead\hfill}               
 \def\@evenhead{\@oddhead}                                              
 \def\@oddfoot{\rm\hfill\thepage\hfill}\def\@evenfoot{\@oddfoot} }      

\makeatother                                                            

\title{Time, measurement and information loss \\ in quantum cosmology}

\def\runninghead{SMOLIN:\quad Time and measurement in quantum cosmology}

\author{
{\em Lee Smolin}
\\Department of Physics, Syracuse University, \\
Syracuse NY USA 13244}
\date{} 

\begin{document}

\pagestyle{headings}                                                   
\flushbottom                                                           
\maketitle
\vspace{-10pt} 

\begin{abstract}
A framework for a physical interpretation of quantum cosmology
appropriate to a nonperturbative hamiltonian formulation is
proposed.
It is based on the use of matter fields to define
a physical reference frame.  In the case of the loop
representation it is
convenient to use a spatial reference frame that picks
out the faces of
a fixed simplicial complex and a clock built with a free scalar
field.  Using these  fields a procedure is proposed for
constructing physical states and
operators in which the problem of constructing physical
operators
reduces to that of integrating ordinary differential equations
within the  algebra of spatially  diffeomorphism invariant operators.
One consequence is that we may conclude that
the spectra of operators that measure the areas of
physical surfaces are discrete independently of the matter
couplings or dynamics of the gravitational field.

Using the physical observables and the physical inner
product,  it becomes possible to describe singularities,
black holes and loss of information in a nonperturbative
formulation of quantum gravity, without making reference
to a background  metric.   While only a dynamical calculation
can answer the
question of whether quantum effects eliminate singularities,
it is conjectured that, if they do not, loss of information is
a likely result because the physical operator algebra that
corresponds to
measurements made at late times must be incomplete.

Finally, I show that it is possible to apply  Bohr's original
operational
interpretation of quantum mechanics to quantum cosmology, so that
one is
free to use either a Copenhagen interpretation or a corresponding
relative
state interpretation in a canonical formulation of quantum
cosmology.

\end{abstract}

\eject
\footnotesize
\tableofcontents
\normalsize
\eject
\section{Introduction}

What happens to the information contained in a star that collapses to
a black hole, { after } that black hole has evaporated?
This question, perhaps more than any other, holds the key to
the problem of quantum gravity.  Certainly, no theory could be called
a successful unification of quantum theory and general relativity that
does not confront it.  Nor does it seem likely that
this can be done without
the introduction of new ideas.    Furthermore, in spite of the progress
that has been made on quantum gravity on several fronts over the
last years, and in spite of some recent attention focused directly
on  it\footnote{An unsystematic sampling
of the  interesting papers that have recently
appeared are
\cite{CGHS,witten-bh,banks-bh,gary-bh,dieter-bh,andy-bh}
\cite{preskill-bh,hawking-2d,lenny-bh,thooft-bh}.}  this problem
remains at this moment  open.

In this paper I would like to ask how this problem may be addressed
from
the point of view of one approach to quantum gravity, which is the
nonperturbative approach based on canonical
quantization\cite{review-abhay,review-carlo,review-lee}.
This approach has been under rapid development for the last
several years in the hopes of developing a theory that could
address such questions from first principles.  What I hope to
show here is that this approach has recently
come  closer to being able
to address problems of physics and cosmology.
To illustrate this, I hope to show here that the
canonical approach
may lead to new perspectives about the problem of what happens
when a black hole evaporates that come from thinking carefully
about how such questions can be asked from a purely diffeomorphic
and nonperturbative point of view.

One reason why the canonical approach has not, so far, had much to
say
about this and other problems is that there is a kind of discipline
that
comes from working completely within a nonperturbative
framework
that, unfortunately,  tends to damp certain kinds of intuitive
or speculative
thinking about physical problems.  This is that, as there is no
background geometry to make reference to,  one cannot
say anything about physics unless it is said using physical operators,
states and inner products.  Unfortunately, while we have gained
some nontrivial
information about the
physical states of the theory, there has been, until recently,
rather little progress about the
problem of constructing physical operators.

{}From a conceptual point of view, the problem of the
physical observables
is difficult because it is closely connected to the problem of
time\footnote{Two good recent reviews of the problem of time
with many original references
are \cite{karel-time,chris-time}.  The point of view pursued here
follows closely that of Rovelli in \cite{carlo-time}.}.  It
is difficult to construct physical observables because in a
diffeomorphism
invariant theory one cannot be naive about where and when an
observation takes place.  Coordinates have
no meaning so that, to be physically meaningful, an
operator must locate the information it is
to measure by reference to the
physical configuration of the system.  Of course, this is not necessary
if we are interested only in global, or topological, information about
the
fields, but as general relativity is a local field theory, with
local degrees
of freedom, and as we are local observers, we must have a
practical way to construct operators that describe local
measurements
if we are to have a useful quantum theory of gravity.

Thus, to return to the opening question, if we are, within
a nonperturbative framework, to ask what happens {\it after a black
hole evaporates,} we must be able to construct spacetime
diffeomorphism
invariant operators that can give physical meaning to the notion of
{\it after the evaporation.}
Perhaps I can put it in the following way:
the questions about loss of information or breakdown of unitary
evolution rely, implicitly, on a notion of time.  Without reference
to time it is impossible to say that something is being lost.
In a quantum theory of
gravity, time is a problematic concept which makes it difficult to
even
ask such questions at the nonperturbative level,
without reference to a fixed spacetime manifold.  The main idea,
which
it is the purpose of this paper to develop, is that the problem of time
in the nonperturbative framework is more than an obstacle that
blocks
any easy approach to the problem of loss of information in black hole
evaporation.  It may be the key to its solution.

As many people have argued, the problem of time is indeed
the conceptual
core of the problem of quantum gravity.  Time, as it is  conceived in
quantum mechanics is a rather different thing than it is from the
point of view of general relativity.  The problem of quantum gravity,
especially when put in the cosmological context, requires for
its solution that some single concept of time be invented that is
compatible with both diffeomorphism invariance and the principle
of superposition.  However, looking beyond this, what is at stake in
quantum gravity is indeed no less and no more than
the entire and ancient mystery:  {\it What
is time?}  For the theory that will emerge from the search for
quantum gravity is likely to be the background for future discussions
about the nature of time, as Newtonian physics
has loomed
over any discussion about time from the seventeenth century
to the present.

I certainly do not know the
solution to the problem of time.  Elsewhere I have
speculated about the direction in which we might search for its
ultimate resolution\cite{napoli}.  In this paper I will
take a rather different
point of view, which is based on a retreat to what both Einstein and
Bohr
taught us to do when the meaning of a physical concept becomes
confused: reach for an operational definition.  Thus, in
this paper I will adopt the point of w that time is precisely
no more and no less than that
which is measured by physical clocks.  From
this point of view, if we want to understand what
time is in quantum gravity then we must construct a description of
a physical clock living inside a relativistic quantum mechanical
universe.

This is, of course, an old idea.  The idea that physically meaningful
observables in general relativity may be constructed by
introducing a physical reference system was introduced  by
Einstein\cite{albert-snail}.  To
my knowledge it was introduced to the literature on
quantum gravity in a classic paper of DeWitt\cite{bryce-snail}
and has recently been
advocated by Rovelli\cite{carlo-matter},
Kuchar and Torre\cite{karel-matter},
Carlip\cite{carlip-matter} and other authors.  However,
what I hope becomes clear from
the following sections
is that this is not just a nice idea which can be illustrated in
simple model systems with a few degrees of freedom.
There is, I believe,
a good chance that this proposal can become the heart
of a viable strategy
to construct physical observables, states  and inner products in the
real
animal-the quantum theory of general relativity coupled to an
arbitrary
set of matter fields.  Whether any of those theories really exist as
good
diffeomorphism invariant quantum field theories is, of course, not
settled
by the construction of an approach to  their interpretation.  However,
what I think emerges from the following is a workable strategy to construct
the theory in a way that, if the construction works, what we will
have in
our hands is a physical theory with a clear interpretation.

The interpretational framework that I  will be proposing is based on
both technical and conceptual developments.  On the technical side,
I will be making use of recent developments that
allow us to construct finite operators that represent diffeomorphism
invariant quantitites\cite{carloobserves,antisymm,viqarobserves}.
These include   spatially
diffeomorphism invariant operators
that measure geometrical quantities such as the
areas of surfaces picked out by the configurations of certain
matter fields.
 By putting this together
with a simple physical
model of a field of synchronized clocks, we will see
that we are able to
implement in full quantum general relativity the
program  of constructing physical observables based
on quantum reference systems .

It may be objected that real clocks and rulers are much more
complicated
things than those that
are modeled here; in reality they consist of multitudes of atoms
held together by electromagnetic interactions.  However, my
goal here is precisely to show that useful results can be
achieved by taking
a shortcut in which the clocks and rulers are idealized and
their dynamics
simplified to the point that their inclusion into the nonperturbative
dynamics is almost trivial.  At the same time, no simplifications
or approximations of any kind are made concerning the dynamics of
the
gravitational degrees of freedom.   What we will then
study is a system in which
toy clocks and rulers interact with the fully nonlinear gravitational
field
within a nonperturbative framework.

However, while I will be using
toy clocks and rulers, the main results
will apply equally to any system in which certain degrees of freedom
can
be used to locate events relative to a physical reference system.  The
chief of these results
is that the construction of physical observables need not be
the very difficult problem that it has  sometimes
been made out to be.  In particular,
it is not necessary to exactly integrate Einstein's equations to find the
observables of the coupled gravity-reference matter system.  Instead,
I propose here an alternative approach which consists of the following
steps:
i)  Construct
a large enough set of spatially diffeomorphism invariant operators to
represent any observations made with the help of a spatial physical
reference system;  ii)  Find the reality conditions among these
spatially diffeomorphism invariant operators, and find the diffeomorphism
invariant inner product that implements them;  iii)  Construct the
projection of the Hamiltonian constraint as a finite and diffeomorphism
invariant operator on this space. iv)  Add  degrees of freedom to
correspond to a clock, or to a field of synchronized clocks.
The Hamiltonian constraint for both states and operators now become
ordinary differential equations for one parameter families of states and
operators in the diffeomorphism invariant Hilbert space parametrized
by the physical time measured by this clock.  v)  Define the physical
inner product
from the diffeomorphism invariant inner product by identifying the
physical inner products of states with the diffeomorphism invariant inner
products of their data at an initial physical time.

The steps necessary to implement this program are
challenging.  But the recent progress, concerning both
spatially diffeomorphism invariant operators
\cite{carloobserves,antisymm,viqarobserves} and the form of
the Hamiltonian constraint
operator\cite{carlolee,berndt-jorge-c,gambinic,milesc}
suggests to me that each step can be accomplished.
If so,  then we will have a systematic way to construct the
physical theory, together with a physical interpretation, from the
spatially diffeomorphism invariant states and operators.

In the next section I review recent results about spatially
diffeomorphism
invariant observables which allow us to implement the idea of a
spatial frame of reference.  In section 3 I show how a simple model
of a field of clocks can be used to promote these to physical
observables\footnote{In an earlier draft of this paper there was
an error in the treatment of the gauge fixed quantization in this
section.  The present treatment corrects the error and is, in addition,
considerably simplified with respect to the original version.}.

Let me then turn from technical developments to  conceptual
developments.  As is well known, there are two kinds of spatial
boundary
conditions that may be imposed in a canonical approach to quantum
gravity: the open and the closed, or cosmological.  The use
of open boundary conditions, such as asymptotic flatness, avoids
some
of the main conceptual issues of quantum gravity because there is a
real
Hamiltonian which is tied to the clock of an observer outside the
system,
at spatial infinity.  However, the asymptotically flat case also
introduces
additional difficulties into the canonical quantization program, so that
it has not, so far, really helped with the construction of the full
theory\footnote{However, there are some interesting developments
along this line, see \cite{baez-asympt}.}.
Furthermore, it can be argued that the asymptotically flat case
represents an idealization that, by breaking the diffeomorphism
invariance
and postulating a classical observer at infinity, avoids exactly those
problems which are the keys to quantum gravity.  Thus, for both
practical and philosophical reasons, it is of interest to see if it is
possible to give a physical interpretation to quantum gravity in the
cosmological context.

There has been a great deal of discussion recently about the
interpretational problems of
quantum cosmology\footnote{See, for example,
\cite{hartle-qc,spain-time,osgood}.}.  However,
most of it is
not directly applicable to the project of this paper, either
because it is tied to the path integral approach to quantum gravity,
because it
is applicable only in the semiclassical limit or because it breaks,
either explicitly
or implicitly,  with the postulate that only operators that
commute with the Hamiltonian constraint can correspond to observable
quantities.   What is required to turn canonical quantum cosmology
into
a physical theory is an interpretation in terms of expectation
values, states
and operators that describes what observers inside the
universe can measure.

At the time I began thinking about this problem it seemed to me
likely that what was required was
some modification of the relative state idea
of Everett\cite{everett}, perhaps
along the lines sketched in \cite{me-everett}, which avoided
commitment to the metaphysical idea of "many worlds" and incorporated
some of the recent advances in understanding of the phenomena of
"decoherence"\cite{spain-time}.
The reason for this was that  it seemed that
the original interpretation of quantum mechanics, as developed by
Bohr, Heisenberg, von Neumann and others could not be applied in the
cosmological context.  However,
I have come to believe that this
is too hasty a conclusion, and that, at least in the context in which
physical observables are constructed by explicit reference to a
physical reference frame and physical clocks,  it is possible to apply
directly to quantum cosmology
the point of view of the original founders of quantum
mechanics.   The key idea is, as Bohr always stressed \cite{Bohr},
to keep throughout the discussion an entirely operational point of
view,
so that the quantum state is never taken as a description of physical
reality but is, instead, part of a description of
a process of preparation and measurement involving
a whole, entangled system including both the quantum system and
the measuring devices.

I want to make it clear from the start that I do not intend
here to take up
the argument about different interpretations of quantum mechanics.
In either ordinary quantum mechanics or in
quantum cosmology there
may be good reasons to prefer another interpretation over the
original
interpretation of Bohr.
What I want to argue here is only that the claim that
it is necessary to give up Bohr and von Neumann's interpretation in
order
to do quantum cosmology is wrong.  As in ordinary quantum
mechanics,
once a strictly operational interpretation such as that of Bohr and von
Neumann has been established, one can replace it with any other
interpretation  that makes more substantive claims about physical
reality, whether it be a relative state interpretation, a statistical
interpretation, or anything else.  For this reason,
I will give, in section 4,
 a sketch
of an interpretation of quantum cosmology  following
the original language of Bohr and von Neumann.  The reader who wants
to augment this with the more substantive
language of Everett, or of
decoherence, will find that they can do so, in quantum cosmology no
more and no less than in
ordinary quantum mechanics\footnote{The question of modeling
the
measurement process in parametrized systems is discussed in
a paper in this volume by Anderson\cite{arley}.  Although
Anderson warns against a too naive application of the projection
postulate that does not take into account the fact
that measurements take a finite amount of time, I do not think
there is any inconsistency between his results and those of the
present paper.}.

Of course, one test that any proposed interpretation of quantum
cosmology
must satisfy is that it give rise to conventional quantum mechanics
and quantum field theory in the appropriate limits.  In section
5  I show how ordinary  quantum field
theory can be recovered by taking limits in which
the gravitational degrees of freedom are treated semiclassically.

Having thus set out both the technical foundations and the
conceptual bases of a physical interpretation of quantum cosmology,
we will then be in a position to see what a fully nonperturbative
approach may be able to contribute to the problems raised by
the existence of singularities and the
evaporation of black holes\cite{hawking-evap}.
While I will certainly not be able to resolve these problems here,
it is possible to make a few preliminary steps that may clarify
how these problems may be treated within a nonperturbative
quantization.  In particular, it is useful to see whether there are
ways in which the existence of singularities and loss of information
or breakdown of quantum coherence could manifest themselves in
a fully nonperturbative treatment that does not make reference
to any classical metric.

What I will show in section 6 is that there are useful notions of
singularity and loss of information that make sense at the
nonperturbative level.  As there is no background metric, these must
be described completely in terms of certain properties of the
physical operator algebra.  The main result of this section is that
this can be done within the context of the physical reference
systems developed in earlier sections.   Furthermore, one
can see at this level a relationship between the two phenomena,
so that it seems likely the existence of certain kinds of singularities
in the physical operator algebra can lead to effects that are
naturally described as "loss of information."   These results indicate
that
the occurrence of singularities and of loss of information are not
necessarily inconsistent with the principles of quantum mechanics
and general relativity.  Whether they actually occur is then a dynamical
question; it is possible that some consistent quantum theories of
gravity allow the existence of singularities and the resulting loss of
information, while others do not.

In order to focus the discussion, the results of section 6
are organized by the
statements of two conjectures, which I call the {\it quantum
singularity conjecture} and the {\it quantum cosmic censorship
conjecture}.  They embody the conditions under which we would want
to say that the full quantum theory of gravity has singularities and
the consequent loss of information.

The concluding section of this paper then focuses on two questions.
First, are
there approaches to an interpretation of quantum cosmology which, not
being based on an operational notion of time, may avoid some of the
limitations of the interpretation proposed here?  Second, are there
models and reductions of quantum cosmology in which the ideas presented
here may be tested in detail?

It is an honor to contribute this paper to a volume in honor
of Dieter Brill, who I have known for 16 years, first as a
teacher of friends, then as a colleague as I became a frequent
visitor to the Maryland relativity group.  I am grateful for the warm
hospitality I have felt from Dieter and the Maryland group on my
many visits there.

\section{A quantum reference system}

In this section I will describe one
example of a quantum reference system in which
relative spatial positions are fixed using the
configurations of certain matter fields.
While I mean for this example to serve as a general paradigm for
how
reference systems  might be described
in quantum cosmology, I will use a coupling to matter and a set of
observables that we have recently learned can be implemented
in nonperturbative quantum
gravity.
Although I do not give the
details here, every operator described in this section may be
constructed by means of a regularization procedure, and
in each case the
result is a finite and diffeomorphism
invariant operator\cite{carloobserves,antisymm,viqarobserves}.

In this section I will speak informally about preparations and
measurements, however the precise statements of the measurement
theory are postponed to section 4; as this depends on the operational
notion of time introduced in the next section.

In this and the following sections, I am describing a canonical
quantization
of general relativity coupled to a set of matter fields.
The spatial manifold,
$\Sigma$, has fixed topological and differential structure, and will be
assumed to be compact.  For definiteness I will make use of the loop
representation formulation of canonical
quantum gravity\cite{abhay,carlolee}.  Introductions
to that formalism are found in
\cite{review-abhay,review-carlo,review-lee}; summaries of results through
the fall of 1992  are found in \cite{abhay-leshouches,cordoba}.

\subsection{Some operators invariant under spatial diffeomorphisms}

In the last year we have found that while it seems impossible to
construct operators that measure the gravitational field at a point,
there exist well defined operators that measure nonlocal observables
such as areas, volumes and parallel transports.  In order to make
these invariant under spatial diffeomorphisms we can   introduce
a set of matter fields which will label sets of open surfaces in
the three manifold $\Sigma$.  I will not here give details of how
this is done, but ask the reader to assume the  existence of
matter fields whose configurations
can be used to label a set of $N$ open surfaces, which I will call
${\cal S}_I$, where $I=1,...,N$.    The boundaries to these surfaces
will also play a role, these are denoted $\partial{\cal S}_I$.

There
are actually
three ways in which such surfaces can be labeled by matter fields.
One can use scalar fields, as described by
Rovelli in \cite{carloobserves} and
Husain in
\cite{viqarobserves}, one can
use antisymmetric tensor gauge fields, as is described
in \cite{antisymm} or one can use abelian gauge fields in
the electric field representation,
as discussed by Ashtekar and Isham \cite{abhaychris}.  In each case we
can construct finite diffeomorphism invariant operators which
measure either the areas of these surfaces or the parallel
transport of the
spacetime connection
around their boundaries.  I refer the reader to
the original papers for the technical details.

The key technical point is that, in
each of these cases,
the matter field can be quantized in a {\it surface representation}, in
which
the states are functionals of a set of $N$ open
surfaces in the three dimensional
spatial
manifold $\Sigma$\cite{gambini-surface,antisymm}.
For each of the $N$ matter fields, a
general bra will then be labeled by an unordered open
surface, which may be disconnected, and will be
denoted $<{S}_I|$ We assume that the states
in the surface representation satisfy an identity which is analogous
to the Abelian loop identities
\cite{gambini-loop,abhay-carlo-maxwell,abhaychris}.  This is that
whenever
 two, possibly disconnected, open  surfaces
${ S}^1$ and ${ S}^2$ satisfy, for every two
form $F_{ab}$,   $\int_{{S}^1} F =\int_{{ S}^2} F $, we
require
that $<{S}_I^1 |= <{ S}_I^2 |$.

A general bra for all $N$ matter fields is then labeled by $N$ such
surfaces, and will be denoted $<{\bf \cal S} |= < {\cal S}_1, ...,{\cal
S}_N |$
so that the general state may be written
\f
\Psi [{\bf \cal S} ] = <{\bf \cal S} |\Psi >
\ff

It is easy to couple this system to general relativity using the loop
representation\cite{carlolee}.  The gravitational degrees of freedom are
incorporated
by labeling the states by both surfaces and loops, so that a general
state is of the form
\f
\Psi [ \gamma , {\bf \cal S}] = <\gamma , {\bf \cal S}|\Psi >
\ff
where $\gamma$ is a loop in the spatial manifold $\Sigma$.
We assume that all of the usual identities of the loop
representation
\cite{carlolee,review-abhay,review-carlo,review-lee} are
satisfied by these states.

The next step is to impose the constraints for spatial
diffeomorphism
invariance.  By following the same steps as in the pure gravity case,
it is easy to see that the exact solution to the diffeomorphism
constraints for the coupled matter-gravity system is that the
states must be functions of the diffeomorphism equivalence classes
of loops and $N$ labeled (and possibly disconnected) surfaces.
As in the case of pure gravity the set of these equivalence classes
is countable.  If we denote by $\{ \gamma , {\bf \cal S}\}$ these
diffeomorphism equivalence classes, every diffeomorphism
invariant
state may be written,
\f
\Psi [ \{ \gamma , {\bf \cal S}\} ] = <\{ \gamma , {\bf \cal S}\}|\Psi >
\ff
Later we will include other matter fields, which will be
denoted generically by $\phi$.  In that case states will be
labeled by diffeomorphism equivalence classes, which I will
denote by
$\{ {\bf \cal S} ,\gamma , \phi \} $.
The space of such states will be denoted ${\cal H}_{diffeo}$.

A word of caution must be said
about the notation: the expression
$<\{ \gamma , {\bf \cal S}\}|\Psi >$ is not to be taken as an
expression of the inner product.  Instead it is just an expression
for the action of the bras states on the kets.  The space of
kets is taken to be the space of functions
$\Psi [ \{ {\bf \cal S} ,\gamma\}  ]$ of diffeomorphism
invariant classes of loops and surfaces.  The space of bras are
defined to be linear maps from this space to
the complex numbers, and given some bra $<\chi |$, the map
is defined by the pairing $<\chi | \Psi >$.  The space of bras
has thus a natural basis which is given by the
$ <\{ \gamma , {\bf \cal S}\}|$, whose action is defined by
(3).  For the moment, the inner product remains unspecified,
so there is no isomorphism between the space of bras and
kets.   At the end of this section I
will give a partial specification of the inner product.

On this space of states it is possible to construct two sets of
diffeomorphism invariant observables
to measure the gravitational field.  The first of
these are the areas of the $I$ surfaces, which I will denote
$\hat{\cal A}^I$.
Operators which measure these areas can
be constructed
in the loop representation.  The details are given in
\cite{carloobserves,antisymm} where it
is
shown that after an appropriate regularization procedure the
resulting quantum operators are diffeomorphism invariant and
finite.

The bras $<\{ \gamma , {\bf \cal S}\}|$ are in fact eigenstates
of the area operators, as long as the loops do not have intersections
exactly at the surfaces.  In references
\cite{carloobserves,antisymm} it is shown that
in this case,
\f
<\{ \gamma , {\bf \cal S}\}|\hat{\cal A}^I = {l_{Planck}^2\over 2 }
{\cal I}^+[{\cal S}^I, \gamma ] <\{ \gamma , {\bf \cal S}\}|
\ff
where ${\cal I}^+[{\cal S}^I, \gamma ] $ is the unoriented, positive,
intersection number between the surface and the loop that simply
counts the intersections between them\footnote{In the case that
there
is an intersection at the surface, the eigenbra's and eigenvalues can
still be found, following the method described in  \cite{review-lee}.}.

A second diffeomorphism invariant observable that can be
constructed is the Wilson loop around the boundary of the $I$'th
surface, which I will denote $\hat{T}^I$.
As shown in \cite{antisymm} it has the action
\f
<\{ \gamma , {\bf \cal S}\}|\hat{T}^I=
<\{ \gamma \cup \partial {\cal S}^I , {\bf \cal S}\}|
\ff
In addition, diffeomorphism
invariant analogues of the higher loop operators have recently
been constructed  by Husain \cite{viqarobserves}.

It is easy to see that the algebra of these operators has the form,
\f
[\hat{T}^J , \hat{\cal A}^I ] = {l_{Planck}^2 \over 2}
{\cal I}^+[{\cal S}^I, \partial {\cal S}^J ] \left (  \hat{T}^J + \
\mbox{intersection \ terms }        \right )
\ff
where intersection terms stands for additional terms that arise if it
happens that the loop $\gamma$ in the quantum state
acted on intersects
the boundary $\partial {\cal S}^J$ exactly
at the surface ${\cal S}^I$ or if
the boundary itself self-intersects at that surface.  We will not need
the detailed form of these terms for the considerations of this paper.

\subsection{Construction of the quantum reference system}

With these results in hand, we may now construct a
quantum reference system.    The
problem
that we must face to construct a measurement theory for
quantum gravity is how to give a diffeomorphism invariant
description of the reference frame and measuring instruments
because, as the geometry  of spacetime is the dynamical variable
we wish to measure, there is no background metric available to
use in their description.  The key idea
is then that a reference frame must  be specified by a particular
topological arrangement of the matter fields that go into its
construction.  In the simple model we are considering here it is very
easy to do this.  As our reference frame is to consist of surfaces,
what we
need to give to specify the reference frame is a particular
topological
arrangement of these surfaces.

One way to do this is the following\cite{antisymm}.  Choose  a
simplicial
decomposition of
the three manifold $\Sigma$ which has $N$ faces, which we may
label
${\cal F}_I$.  For reasons that will be clear in a moment, it is
simplest
to restrict this choice to simplicial decompositions in which the
number of
edges is also equal to $N$.  Let me call such a choice ${\cal T}$.

Now, for each ${\cal T}$ with $N$ faces there is a subspace of the
state space ${\cal H}_{diffeo}$ which is spanned by basis elements
$<\{ \gamma , {\bf \cal S}\}|$ in
which the surfaces ${\cal S}_I$ can be put into a
one to one correspondence
with the faces ${\cal F}_I$ of ${\cal T}$ so  that they have the
same
topology of the faces of the simplex.    We may call this subspace
${\cal H}_{{\cal T}, diffeo}$.

As I will describe in section 4, the preparation of the system is described
by putting the system into such a subspace of the Hilbert space associated
with an arrangement of the surfaces.
Once we know the state is in the subspace
${\cal H}_{{\cal T}, diffeo}$
any measurements of the quantities
$\hat{\cal A}^I$ or $\hat{T}^I$
can be interpreted in terms of areas of the faces ${\cal F}^I$
or parallel transports around their edges.

\subsection{How do we describe the results of the measurements?}

Given a choice of the
simplicial manifold $\cal T$ and the corresponding subspace
${\cal H}_{{\cal T}, diffeo}$
we may now make measurements of the gravitational field.  As I
will establish in section 4, there will be circumstances
in
which it is meaningful to say that we have, at some particular time,
made a measurement of some commuting subset of the operators
$\hat{\cal A}^I$ and $\hat{T}^I$ which we described above.
We may first note that these operators are block diagonal in
${\cal H}_{diffeo}$ in that their action preserves the subspaces
${\cal H}_{{\cal T}, diffeo}$.
{}From the commutation relations (6) we may deduce that if we
restrict attention to one  of these subspaces and these
observables there are two
maximal sets of commuting operators; we may measure either the
$N$
$\hat{\cal A}^I$ or the $N$ $\hat{T}^I$.  This corresponds directly to
the
fact that the canonical pair of fields in the Ashtekar formalism
are the spatial frame field and the self-dual connection.

How are we to describe the results of these measurements?  Each
gives us $N$ numbers, which comprise partial information that
we can obtain about the geometry of spacetime as a result of a
measurement
based on the quantum reference frame built on ${\cal T}$.  Now, as
in ordinary quantum mechanics, we would like to construct a
classical
description of the result of such a partial measurement of the system.
I will now show that in each case it is possible to do this.  What we
must
do, in each case, is associate to the results of the measurements, a
set of
classical gravitational fields that are described in an appropriate
way
by $N$ parameters.

It is simplest to start with the measurements of the areas, which
give
us a partial measurement of the spatial geometry.

\subsection*{Classical description of the output of the
measurements of the areas }

The output of a measurement of  the $N$ areas will be $N$ rational
 numbers,
$a_I$, (times the Planck area),
each from the discrete series of possible eigenvalues
in the spectra of the $\hat{A}^I$.  Let us associate to each such
set of areas a piecewise flat three geometry  ${\cal Q}(a)$  that can
be constructed as follows.  ${\cal Q}(a)$ is the Regge
manifold constructed by putting together
flat tetrahedra according to the topology given by ${\cal T}$ such
that
the areas of the $N$ faces ${\cal F}_I$ are given by
$a_I l_{Planck}^2$.
Since such a Regge manifold is defined by its edge lengths
and since we have fixed ${\cal T}$ so that the number of its edges is
equal to the number of its faces, the $N$ areas $a_Il_{Planck}^2$
will
generically determine the $N$ edge lengths.

Note that we are beginning with the assumption that all of the
areas are positive real numbers, so the triangle inequalities must
always be satisfied by the edge lengths.  At the same time, the
tetrahedral identities may not be satisfied, in that there is
no inequality which restricts the areas of the faces of
an individual tetrahedron in $\cal T$.  For example, there exist
configurations $\{ \gamma , {\bf \cal S} \}$ in which the loop
only intersects one of the faces, giving that one face a finite
area while the remaining faces have vanishing area.   Thus, we
must include the possibility that ${\cal Q}(a)$ contains tetrahedra
with  flat metrics with indefinite signatures.
The emergence of a geometry that, at least when measured on large
scales, may be approximated by a positive definite metric must
be a property of the classical limit of the theory.

There may also be special cases
in which more than one set of edge lengths are consistent with the
areas.  In which case we may say that the measurement of the
quantum geometry leaves us with a finite set of possible classical
geometries.  There is nothing particularly troubling about this,
especially as this will not be the generic case.

Thus, in general, the outcome of each measurement of the
$N$ area operators may be describe by a particular piecewise
flat Regge manifold, which represents the partial measurement
that has been made of the spatial geometry.

\subsection*{Classical description of the output of the
measurements
of the self-dual parallel transports}

What if we measure instead the $N$ Wilson loops,  $\hat{T}^I$?
The output of such a measurement
will be $N$ complex numbers, $t^I$.
Can we associate these with a classical construction?  I want to
show
here that the answer is yes.

Any such classical construction must not involve a spatial metric,
as
we have made the spatial geometry uncertain by measuring the
quantities conjugate to it.  So it must be a construction which is
determined by $N$ pieces of information about the self-dual
part of the spacetime curvature.

Such a construction can be given, as follows.  We may construct a
dual
graph to ${\cal T}$ in the natural way
by associating to each of its tetrahedra a vertex and to each
of its faces, ${\cal F}_I$ an edge, called $\alpha_I$, such that the
$4$ $\alpha_I$ associated with the faces of a given tetrahedra have
one of their  end points at the vertex that corresponds to it.
As each face
is part of two tetrahedra, we know
where to put the two end points of each edge, so the
construction is completely determined.
We may call this dual graph $\Gamma_{\cal T}$.

Now, to each such graph we may associate a distributional self-dual
curvature which is written as follows,
\f
F_{ab}^i (x) = \sum_I \int d\alpha^c_I (s) \ \epsilon_{abc} \
\delta^3 (x , \alpha_I(s) ) \  b^i_I
\ff
which is determined by giving $N$ $SL(2,C)$ algebra elements
$b^i_I$.
If we use the non-abelian Stokes theorem \cite{nastokes},
we may show that
the moduli of the $N$ complex $b^i_I$ are determined
by the $N$
complex
numbers $t_I$ that were the output of the measurement by
\f
{1 \over 2} Tre^{\imath b^i_I\tau^i } = cos |b_I| = t_I
\ff
where $\tau^i$ are, of course, the three Pauli matrices.  The
remaining
information about the orientation of the $b^i_I$ is gauge dependent
and
is thus not fixed by the measurement.

The reader may wonder whether a connection field can be associated
with
a distributional curvature of the form of (7).
The answer is yes, what
is
required is a Chern-Simon connection
with source given by (7).  For
any such source there are solutions
to the Chern-Simon equations,
which, however, require additional structure to be fully specified.
One
particularly simple way to do it, which does not depend on the
imposition
of a background metric, is the
following\cite{review-lee,newlattice}.  Let us give an
arbitrary  specification of
the faces of the dual graph ${\cal G}_{\cal T}$, which I will call
${\cal K}_I$.  We may note that there is one face of the dual graph
for each edge of ${\cal T}$, so that their number is also equal
to $N$.  Then we may specify a distributional connection of the form,
\f
A_a^i (x) = \sum_I \int d^2 {\cal K}_I^{bc} (\sigma ) \epsilon_{abc}
\delta^3 (x, {\cal K}_I(\sigma ) ) a_I^i
\ff
where the $a_I^i$ are, again, $N$ Lie algebra elements.  It is then
not difficult to show that the usual relationship between the
connection
and the curvature holds, in spite of their distributional form and
that
the  $b_I^i$'s may be expressed in terms of the $a_I^i$'s.  For details
of this the reader is referred to \cite{newlattice}.

Thus, we have shown that to each measurement of the $N$ Wilson
loops of the self-dual connection, we can also associate a classical
geometry, whose construction is determined by $N$ pieces of
information (in this case complex numbers.)  The result may
be thought of as a partial determination of the geometry of a
spacetime Regge manifold.  If we add a dimension, and allow time
to be discrete than the construction we have just given can be
thought
of as a spatial slice through a four dimensional Regge manifold.  In
that case, each edge in our construction becomes a face in the four
dimensional construction and, as in the Regge case, the curvature is
seen to be distributional with support on the faces.  However, in this
case it is only the self-dual part of the curvature that is given,
because
its measurement makes impossible the measurement of any
conjugate
information.  In fact, a complete construction of a Regge-like four
geometry can be given along these lines, for
details, see \cite{newlattice}.

\subsection{The spatially diffeomorphism invariant inner product}

In all of the constructions so far given, the inner product has played
no
role because we have been expressing everything in terms of
the eigenbras of the operators in a particular basis, which are the
$<\{ \gamma , {\bf \cal S}\}|$.  However, as in ordinary quantum
mechanics, a complete description of the measurement theory will
require an inner product.   The complete specification of the inner
product must be done at the level of the physical states, which
requires that we take into account that the states are solutions
to the Hamiltonian constraint.  This problem, which I would like to
claim is essentially equivalent to the problem of time, is the
subject
of the next section.  But it is interesting and,
 as we shall see, useful to see how
much can be determined about the inner product at the level of
spatially diffeomorphism invariant states.

Now, in order to determine the
inner product at the diffeomorphism invariant level we should be
able to write the reality conditions that our classical
diffeomorphism
invariant observables satisfy.  For the $\hat{T}^I$ this is an unsolved
problem, these operators are complex, but must satisfy reality
conditions which are determined by the reality conditions on the
Ashtekar connections.  To solve this problem it will be necessary
to adjoin additional diffeomorphism invariant operators to the
$\hat{T}^I$ and $\hat{\cal A}^I$ in order to enable us to write down
a complete star algebra of diffeomorphism invariant observables.

However, the reality conditions the area operators satisfy are very
simple:
they must be real.  As a result we may ask what restrictions we may
put on the inner product such that
\f
\hat{\cal A}^{I \dagger }=\hat{\cal A}^{I}  ?
\ff
To express this we must introduce characteristic kets, which
will be denoted
$|\{ \gamma , {\bf \cal S}\} >$.  They are  defined so that
\f
\Psi_{\{ \gamma , {\bf \cal S}\} }
[\{ \gamma , {\bf \cal S}\}^\prime] \equiv
<\{ \gamma , {\bf \cal S}\}^\prime |\{ \gamma , {\bf \cal S}\}> =
\delta_{\{ \gamma , {\bf \cal S}\}\{ \gamma , {\bf \cal S}\}^\prime}
\ff
Here the meaning of the delta is follows.  Fix a particular, but
arbitrary
set of surfaces and loops $( \gamma , {\bf \cal S}) $ within the
diffeomorphism
equivalence class $\{ \gamma , {\bf \cal S}\}$.  Then the
$\delta_{\{ \gamma , {\bf \cal S}\}\{ \gamma , {\bf \cal S}
\}^\prime}$
is equal to one if and only if there is an element ${\bf \cal S}^\prime
,\gamma^\prime $ of the equivalence class $\{ \gamma , {\bf \cal S}
\}^\prime$ such that a) for
every two form $F_{ab}$ on $\Sigma $,
$\int_{{\cal S}^\prime}F=\int_{{\cal S}}F$ and b) for every
connection
$A_a^i$ on $\Sigma $ and every component $\gamma_{\cal I}$ of
$\gamma$ (and similarly for $\gamma^{\prime}$),
$T[\gamma_{\cal I}^\prime ]=T[\gamma_{\cal I}]$.   If the
condition is not satisfied then
$\delta_{\{ \gamma , {\bf \cal S}\}\{ \gamma , {\bf \cal S}
\}^\prime}$
is equal to zero.

Let me denote the
diffeomorphism invariant
inner product by specifying the adjoint  map
from kets to bras,
\f
<\{ \gamma , {\bf \cal S}\}^\dagger |
\equiv |\{ \gamma , {\bf \cal S}\}>^\dagger.
\ff
It is then straightforward to show that the
condition (10) that the $N$ operators must be hermitian restricts
the inner product so that, in the case that the loops $\gamma$
have no intersections with each other,
\f
<\{ \gamma , {\bf \cal S}\}^\dagger | =<\{ \gamma , {\bf \cal S}\}|.
\ff

\section{Physical observables and the problem of time}

In this section I would like to describe an operational approach
to the problem of time in quantum cosmology, which is
based on the point of view that time is no more and no less than
that which is measured by physical clocks.
The general idea we will pursue is  to
couple general relativity to a matter field
whose behavior makes it suitable for use as a clock.    One
then turns the Hamiltonian constraint equations into evolution equations
that proscribe how spatially diffeomorphism invariant quantities
evolve
according to the time measured by this physical clock.

We will then try to build up
the physical theory with the clock from the
spatially diffeomorphism invariant theory for the gravitational and
matter
degrees of freedom,  in which the clock has been left out.
To do this
we will need  to assume  several things about
the diffeomorphism
invariant theory, which are motivated by the results of the last
section.

a)  In the loop representation we have the complete set of solutions
to
the spatial diffeomorphism constraints coupled to more or less
arbitrary matter fields.  Given some choice of matter fields, which I
will denote generically by $\phi$, I will write the general spatially
diffeomorphism invariant state by
$\Psi[\{ \gamma , \phi \}]_{diffeo}$,
where the brackets $\{...\}$ mean spatial diffeomorphism
equivalence
class.  The reference frame fields discussed in the previous section
are, for the purpose of simplifying the formulas of this
section, included in the $\phi$.  However, the fields that
represent the clock {\it are not} to be included in these
diffeomophism invariant states.  As in the
previous section,   the Hilbert space
of diffeomorphism invariant states will
be denoted ${\cal H}_{diffeo}$.

b)  I will assume that an algebra of finite
and diffeomorphism invariant operators, called ${\cal A}_{diffeo}$,
is known on ${\cal H}_{diffeo}$.  The idea that spatially
diffeomorphism
invariant operators are finite has become common in the loop
representation, where there is some evidence for the conjecture
that diffeomorphism invariance requires finiteness
\cite{weaves,carloobserves,antisymm,review-lee}.  The
area and parallel transport operators
discussed in the previous section are examples of such operators.  I
will
use the notation $\hat{\cal O}^I_{diffeo}$ to refer to elements of
 ${\cal A}_{diffeo}$, where $I$ is an arbitrary index that labels
the operators.

c)  I will assume also that the classical diffeomorphism invariant
observables ${\cal O}^I_{diffeo}$ that correspond to the quantum
$\hat{\cal O}^I_{diffeo}$ are known.
This lets us impose the reality conditions
on the algebra, as described in
\cite{review-abhay,abhaychris,review-carlo}.

d)  Finally,
I assume that the inner product $<|>_{diffeo}$ on ${\cal H}_{diffeo}$
has been determined from the reality conditions satisfied by
an appropriate subset of the $\hat{O}^I_{diffeo}$.

\subsection{A classical model of a clock}

I will now introduce a new scalar field, whose value will be
defined to {\it be} time.  It will be called the
clock field and written $T(x)$, and it will be assumed to
have the unconventional dimensions of time.
Conjugate to it we must have a density, which has dimensions
of energy density,which will be
denoted $\tilde{\cal E}(x)$ such that\footnote{We adopt the
convention
that the delta function is a density of weight one on its first entry.
Densities will usually, but not
always, be denoted by tildes.},
\f
\{ \tilde{\cal E}(x) , T(y)  \} = -\delta^3 (x,y)  .
\ff

To couple these fields to gravity we must add appropriate terms
to the diffeomorphism and Hamiltonian constraints.  I will
assume that $T(x)$ is a  free massless
scalar field, so that
\f
{\cal C} (x) = {1 \over 2 \mu} \tilde{\cal E}^2 +
{\mu \over 2}  \tilde{\tilde{q}}^{ab}
\partial_a T \partial_b T  + {\cal C}_{grav} (x)+ {\cal C}_{matter}(x)   ,
\ff
where ${\cal C}_{grav}(x)$ and $ {\cal C}_{matter}(x)$
are, respectively,
the contributions to the Hamiltonian
constraint
for the gravitational field and the other matter fields and
$\mu$
is a  constant with dimensions of {\it energy density}.
Note that the form of (16) is dictated
by
the fact that in the Ashtekar formalism the Hamiltonian constraint
is a density of weight two.

Similarly, the diffeomorphism constraint becomes
\f
{\cal D}(v) =
\int_{\Sigma}v^a (\partial_a T)\tilde{\cal E}
+{\cal D}_{grav}(v)+{\cal D}_{matter}(v)
\ff
 The reader may check that these constraints close in the proper
way.

We may note that because there is no potential term for the clock
field
we have a constants of motion,
\f
 {\cal E} \equiv \int_\Sigma  d^3x \tilde{\cal E}(x).
\ff
This generates the symmetry $  T(x)\rightarrow T(x)+
\mbox{constant}$.
It is easy to verify explicitly that
\f
\{{\cal C} (N) ,  {\cal E}  \}=
\{ {\cal D}  (v) ,  {\cal E} \}=0 ,
\ff
(where ${\cal C} (N) \equiv \int N {\cal C}$) .

We would now like to chose a gauge in which the time slicing of
the spacetime is made according to surfaces of constant $T$.  We
thus choose as a gauge condition.
\f
\partial_a T (x) =0 .
\ff
This may be solved by setting $T(x)=\tau$, where $\tau$ will
be taken to be the time parameter.
We may note that
the condition that the evolution follows surfaces of constant
$T(x)$ fixes the lapse because,
\f
\dot{ (\partial_a T)} (x) = \{ {\cal C}(N) , \partial_a T(x) \}
= \partial_a (N(x) \tilde{\cal E}(x)) =0
\ff
Thus, our gauge condition can only be
maintained if
\f
N(x) = {c \over \tilde{\cal E} (x) }  .
\ff
where $c$ is an arbitrary constant.

One way to say what this means is that all but one of the infinite
number of Hamiltonian constraints have been broken by imposing
the gauge condition (19).   The one remaining component
of the Hamiltonian constraint is the one that satisfies (21).
However, since we must have eliminated the nonconstant
piece of $T(x)$ by the gauge fixing (19) we must solve
the constraints which have been so  broken to eliminate
the fields which are conjugate to them.
Thus, all but one
of the degrees of freedom in $\tilde{\cal E}(x)$ must be eliminated
by solving the Hamiltonian constraint.  The one which is
kept independent can be taken to be the global constant
of motion $\cal E$ defined by (17).  Up to this one overall
degree of freedom, the local variations in ${\cal E}(x)$ must
be fixed by solving the Hamiltonian constraint locally, which
gives us\footnote{We make here a choice in taking the positive
square root, which is to restrict attention to a subspace
of the original  phase
space.  This choice, which we carry through as well
in the quantum theory,
is the equivalent of a positive frequency condition.},
\f
\tilde{\cal E}(x) = \sqrt{-2\mu [ {\cal C}_{grav} (x)
+ {\cal C}_{matter}(x)  ]}
\ff

Note that, to keep the global quantity ${\cal E}$ independent,
we should solve this equation at all but one,
arbitrary,  point of $\Sigma$.

The idea is then to reduce the phase space and constraints
so that the local variations in $T(x)$ and $\tilde{\cal E}(x)$
are eliminated,  leaving only the global variables $\tau$
and ${\cal E}$.
We may  note that
\f
\{ T(x) , {\cal E} \} =1 .
\ff
so that the reduced Poisson bracket structure must give
\f
\{ \tau , {\cal E} \} =1
\ff
while the Poisson brackets of the gravitational and matter fields
remain as before.
After the reduction, we then
have one remaining Hamiltonian constraint,
which is
\begin{eqnarray}
{\cal C}_{g.f.}={\cal C}(c/\tilde{E}) &=&{c \over 2\mu }{\cal E} +
c \int { d^3x \over {\cal E}(x)}
\left ( {\cal C}_{grav} (x) + {\cal C}_{matter}(x) \right )
\nonumber \\
&=& {c \over 2\mu }{\cal E} +
{c \over \sqrt{2 \mu}} \int d^3x \sqrt{-  [ {\cal C}_{grav} (x)
+ {\cal C}_{matter}(x) ] }
\end{eqnarray}

By (24) ${\cal C}_{g.f.}$
 generates reparametrization of the global time variable
$\tau$.
The effect of the reduction on the diffeomorphism constraint
is simply to eliminate the time field so that
\f
{\cal D}(v)_{g.f.} =
{\cal D}_{grav}(v)+{\cal D}_{matter}(v)
\ff
We may note that the reduced Hamiltonian constraint is invariant
under diffeomorphisms and hence commutes with the reduced
diffeomorphism constraint.

To summarize, the gauge fixed theory is based on a phase space
which consists of the original gravitational and matter phase
space, to which the two conjugate degrees of freedom
$\tau$ and ${\cal E}$ have been added.  The diffeomorphism
constraint remains the original one while there remains
one Hamiltonian constraint given by (25).

It is now straightforward to construct physical operators.
They must be of the form
\f
{\cal O}={\cal O}[A,E,\phi, \tau, {\cal E}]
\ff
where $A,E$ are the canonical variables that describe the
gravitational
field and $\phi$ stands for any other matter fields.
The diffeomorphism constraints
imply that
\f
{\cal O} [A,E,\phi, \tau, {\cal E} ]={\cal O} [\{ A,E,\phi \} , {\cal E},
\tau ]
\ff
where the brackets $\{...\}$ indicate that the observable can depend
only
on combinations of the gravitational and matter fields that are
spatially
diffeomorphism invariant.  The requirement that the
observable commute with the reduced Hamiltonian constraint
gives us
\f
{d{\cal O} [A,E,\phi, \tau, {\cal E}]
\over d\tau }= \sqrt{2\mu}
\left \{  \int d^3x \sqrt{-  [{\cal C}_{grav} (x)
+ {\cal C}_{matter}(x) ] },
{\cal O} [A,E,\phi, \tau, {\cal E}] \right \} .
\ff
Thus, we have achieved the following result concerning
physical observables:

\blankline
\noindent
{\it For every spatially diffeomorphism invariant observable
${\cal O} [\{ A,E,\phi \} ]_{diffeo}$ which is a function of the
gravitational
and matter fields (but not the clock degrees of freedom)  there is a
physical observable whose expression in
the gauge given by (19) and (21) is the two parameter family of
diffeomorphism
invariant observables\footnote{The subscript $g.f.$ will be
used throughout this paper
to refer to observables that commute with the full spatial
diffeomorphism constraints but only the gauge fixed
Hamiltonian constraint.},
of the form
${\cal O} [\{ A,E,\phi \} , \tau, {\cal E} ]_{g.f.}$  which solves
(29)  subject to the initial condition that}
\f
{\cal O} [\{ A,E,\phi \} , \tau = 0 ,  {\cal E}]_{g.f.}
= {\cal O} [\{ A,E,\phi \}]_{diffeo}
\ff

By construction, we may conclude that the  observable
${\cal O} [\{ A,E,\phi \} , \tau ]_{g.f.}$ is the value of
the diffeomorphism invariant function
${\cal O} [\{ A,E,\phi \}]_{diffeo}$ evaluated on the surface
$T(x)=\tau  $ of the spacetime gotten by evolving
the constrained initial data $\{ A,E,\phi , T=0\}$.

Now, ${\cal O} [\{ A,E,\phi \} , \tau, {\cal E}]_{g.f.}$ is the
value of a physical observable only in the gauge picked by
(19) and (21)).  However,
once we know the value of any observable in a fixed gauge we
may extend it to a fully gauge invariant observable.  To do so we
look for
a gauge invariant function\footnote{The subscript $Dirac$ will
always refer to an observable that commutes with the full
set of constraints without gauge fixing.}
${\cal O}[A,E,\phi, T(x), {\cal E}(x)]_{Dirac}$
 that commutes
with the full diffeomorphism and Hamiltonian constraints,
with arbitrary lapses $N$, that has the same physical
interpretation as our gauge fixed observable
${\cal O} [\{ A,E,\phi \} , \tau , {\cal E}]_{g.f.}$.  This
means that
\f
{\cal O}
[A,E,\phi, T(x) ,
\tilde{\cal E}(x) ]_{Dirac}
|_{T(x)=\tau ,
\tilde{\cal E}(x)=\sqrt{-2\mu [ {\cal C}_{grav} (x)
+ {\cal C}_{matter}(x)  ]}}
= {\cal O} [A,E,\phi, \tau, {\cal E} ]_{g.f.}.
\ff

Once we have one such physical observable, we may follow Rovelli
\cite{carlo-time,carlo-matter} and
construct a one parameter family of physical observables
called "evolving constants of motion".  These are fully gauge invariant
functions on the phase space that, for each $\tau$ tell us
the value of the spatially diffeomorphism invariant observable
${\cal O} [\{ A,E,\phi \}]_{diffeo}$ on the surface
$T(x)=  \tau$
as a function of the data of the initial surface.
That, is for each $\tau$ we seek a function
${\cal O}^{\prime} [A,E,\phi, T(x),{\cal E}(x) ](\tau) $
that commutes with the Hamiltonian constraint for all $N$ (with
$\tau$ taken as a parameter and not a function on the phase
space) and which
satisfies
\f
{\cal O}^{\prime}
[A,E,\phi, T(x), \tilde{\cal E}(x) ](\tau )
|_{T(x)=0,
\tilde{\cal E}(x)=\sqrt{-2\mu [ {\cal C}_{grav} (x)
+ {\cal C}_{matter}(x)  ]}}
= {\cal O} [A,E,\phi, \tau, {\cal E} ]_{g.f.}  .
\ff

\subsection{Quantization of the theory with the time field}

We would now like to extend the result of the previous section to
the quantum theory.   To do this we must introduce an appropriate
representation for the clock fields and construct and
impose the diffeomorphism and  Hamiltonian constraint equations.

We will first construct the quantum theory corresponding
to the reduced classical dynamics that follows from the gauge
fixing (19) and (21).
After this we will discuss the alternative possibility, which is
to construct the physical theory through Dirac quantization in which
no gauge fixing is done.

In the gauge fixed quantization the states
will be taken to be functions
$\Psi [\gamma , \phi, \tau]$ so that
\f
\hat{\tau} \Psi [\gamma , \phi, \tau]=
\tau \Psi [\gamma , \phi, \tau],
\ff
\f
\hat{\cal E} \Psi [\gamma , \phi, \tau]=
-\imath \hbar {\partial \over  \partial \tau}
\Psi [\gamma , \phi, \tau]
\ff
and all the other defining relations are kept.  The space of these
states, prior to the imposition of the remaining
constraints, will be called ${\cal H}_{reduced}$.

We now apply the reduced diffeomorphism constraints (26).
The result is that the states be
functions of diffeomorphism equivalence classes of
their arguments, so
that
\f
{\cal D}(v)_{g.f.}  \Psi [\gamma , \phi, \tau]=0 \Rightarrow
 \Psi [\gamma , \phi, \tau]=
\Psi [\{ \gamma , \phi \},  \tau]
\ff
where, again, the brackets indicate diffeomorphism
equivalence classes.

We may then apply the reduced Hamiltonian constraint (25).
By (34) this implies that, formally
\f
{\imath \hbar \partial \Psi [\{ \gamma , \phi \} , \tau]
\over \partial \tau}     =
\int d^3x \sqrt{- 2\mu [ \hat{\cal C}_{grav} (x)
+ \hat{\cal C}_{matter}(x)  ] }   \Psi [\{ \gamma , \phi \} , \tau]
\ff

As this is the fundamental equation of the quantum theory, we
must make some comments on its form.  First, and most
importantly, as the Hamiltonian constraint involves operator
products, this equation must be defined through a suitable
regularization procedure.  Secondly, to make sense of this equation
requires that we define an operator square root.   Both steps must
be done in such a way that the result is a finite and diffeomorphism
invariant operator.  That is, for this approach to quantization
to work, it must be possible to regulate the gravitational
and matter parts of the Hamiltonian constraint in such a way that
the limit
\f
\lim_{\epsilon \rightarrow 0}
\int d^3x \sqrt{- 2\mu [ \hat{\cal C}_{grav}^\epsilon (x)
+ \hat{\cal C}_{matter}^\epsilon (x)  ] }  = \hat{\cal W}
\ff
(where the $\epsilon$'s denote the regulated operators) exists
and gives a well defined (and hence finite and diffeomorphism
invariant) operator
$\hat{\cal W}$ on the space of spatially diffeomorphism invariant
states of the gravitational and matter fields.

It may seem that to ask that it be possible to
both define a good regularization procedure and define
the operator square root is to be in danger of being ruled out
by the "two miracle" rule: it is acceptable practice in theoretical
physics to look forward to the occurrence
of one miracle, but to ask for two is unreasonable.  However,
recent experience with constructing diffeomorphism invariant
operators in the loop representation of quantum gravity suggests
the opposite: these two problems may be, in fact, each others
solution.  Perhaps surprisingly,
what has been found is that the only
operators which have been so far constructed as finite,
diffeomorphism invariant operators involve operator square roots.

The reason for this is straightforward.  In quantum field theory
operators are distributions.  In the context of diffeomorphism
invariant theories, distributions are densities.  As a result, there
is an intrinsic difficulty with defining operator products in a
diffeomorphism invariant theory through a renormalization
procedure.  Any such procedure must give a way to define
the product
of two distributions, with the result being another distribution.  But
this means that the procedure must take a geometrical object which
is formally a density of weight two, and return a density of weight
one.  What happens as a result is that there is a grave risk of
the regularization procedure breaking the invariance under
spatial diffeomorphism invariance, because the missing density
weight ends up being represented by functions of the
unphysical background used in the definition of the renormalization
procedure.  Many examples are known in
which exactly this happens\cite{review-lee}.

However, it turns out that in many cases the square root of the
product of two distributions can be defined as another distribution
without ambiguity due to this problem of matching density weights.
It is, indeed, exactly this fact that makes it possible to define the
area operators I described in the previous
section, as well as other operators associated
with volumes of regions and norms
of one form fields\cite{weaves,review-lee}.

I do not know whether the same procedures that work in the
other cases work to make the limit (37) exist.  We may note,
parenthetically,
that if $\hat{\cal W}$ can be defined as an operator
on diffeomorphism invariant states
in the context of a separable Hilbert
structure, the problem of the finiteness of quantum
gravity will have been solved.  I will assume here
that the problem of constructing a regularization
procedure such that this is the case
can be solved, and go on.

Assuming then the existence of $\hat{\cal W}$, we may call
the space of solutions to (35) and (36) ${\cal H}_{g.f}$, where
the subscript denotes again that we are working with the
gauge fixed quantization.  We will shortly be discussing the
inner product on this space.  For the present, the reader may
note that
once $\hat{\cal W}$ exists, the problem of finding states
that solve the reduced Hamiltonian constraint is essentially
a problem of ordinary quantum mechanics.  For example,
there will be solutions of the form
 \f
 \Psi [\{ \gamma , \phi \} ,\tau] =
 \Phi  [\{ \gamma , \phi \}]
e^{-\imath \omega\tau }
\ff
This will solve (36) if $\Phi [\{ \gamma , \phi \} ]$
is an eigenstate
of $\hat{\cal W}$, so that
\f
\hat{\cal W} \Phi = \hbar \omega\Phi
\ff

More generally, given any diffeomorphism invariant
state $\Psi [\{ \gamma , \phi \}] \in {\cal H}_{diffeo}$,
which is a function of the gravitational and matter fields
alone, there is a physical state,
$\Psi [\{ \gamma , \phi \}, \tau ]  \in {\cal H}_{g.f}$
in our gauge fixed quantization
which is the solution to (36) with the initial conditions
\f
\Psi [\{ \gamma , \phi \}, \tau ]_{\tau=0} =
\Psi [\{ \gamma , \phi \}]
\ff
Thus
what we have established is that there is  a map
\f
\Lambda: {\cal H}_{diffeo} \rightarrow {\cal H}_{g.f.}
\ff
in which every diffeomorphism invariant state of the
gravitational and matter fields is taken into its evolution
in terms of the clock fields.
Furthermore, there is an inverse map,
\f
\Theta : {\cal H}_{g.f.}\rightarrow {\cal H}_{diffeo}
\ff
which is defined by evaluating the physical state at
$ \tau=0 $.

\subsection{The operators of the gauge fixed theory}

The physical operators in the gauge fixed quantum
theory may be found analogously to the classical
observables of the gauge fixed theory.  A physical
operator is an operator on ${\cal H}_{g.f.}$, which
may be written
$\hat{\cal O} [\hat A,\hat E,\hat \phi, \hat \tau, \hat {\cal E}]$.
The requirement that it commute with the reduced
diffeomorphism constraints restricts it to be of the
form
$\hat{\cal O} [\{ \hat A,\hat E,\hat \phi \} , \hat \tau, \hat {\cal E}]$.
The requirement that it commute with the reduced Hamiltonian
constraint becomes the evolution equation,
\f
\imath \hbar {d\hat {\cal O}
[\hat A,\hat E,\hat \phi, \tau, \hat {\cal E}]
\over d\tau }=
\left [  \hat{\cal W},
\hat {\cal O} [\hat A,\hat E,\hat \phi, \tau, \hat {\cal E}] \right ] .
\ff

Using this equation we may find a physical operator that
corresponds to every spatially diffeomorphism invariant
operator on ${\cal H}_{diffeo}$, which depends only on
the non-clock fields.  Given any such operator,
$\hat {\cal O} [\{ \hat A,\hat E,\hat \phi\} ]_{diffeo}$
we may construct an operator on ${\cal H}_{g.f.}$ which
we denote $\hat {\cal O} [\{ \hat A,\hat E,\hat \phi\}, \tau ]_{g.f.}$
which solves (43) subject to the initial condition
\f
\hat {\cal O} [\{ \hat A,\hat E,\hat \phi\}, \tau =0  ]_{g.f.}
= \hat {\cal O} [\{ \hat A,\hat E,\hat \phi\} ]_{diffeo}
\ff
We may, indeed, solve (43) to find that
\f
\hat {\cal O} [\{ \hat A,\hat E,\hat \phi\}, \tau  ]_{g.f.}
= e^{-\imath \hat{\cal W}\tau/\hbar }
\hat {\cal O} [\{ \hat A,\hat E,\hat \phi\}  ]_{diffeo}
e^{+\imath \hat{\cal W}\tau / \hbar }  .
\ff

\subsection{The physical interpretation and inner product of the
gauge fixed theory}

In the classical theory, the physical
observables ${\cal O} [\{  A, E, \phi\}, \tau  ]_{g.f.}$ were
found to correspond to the values of diffeomorphism invariant
functions of the non-clock fields on the surface defined by
the gauge condition (19) and
the value of the time parameter $\tau$.  As they satisfy the
analogous quantum equations we would like to
interpret the corresponding
quantum operators that solve (43) and (44)
in the same way.  That is, we will take
$\hat{\cal O}[\{ \hat A,\hat E,\hat \phi\}, \tau  ]_{g.f.}$
 to be the operator that measures
the diffeomorphism invariant quantity
$\hat{\cal O}[\hat{E},\hat{A},\hat{\phi}]_{diff}$ after a physical
time $\tau$.

Once this interpretation is fixed, there is a natural choice for the
physical inner product.  The idea,
 advocated by Ashtekar \cite{review-abhay}, is that the
physical inner product is to be picked to satisfy the reality conditions
for a large enough set of physical observables.  The difficult part
of the definition is the meaning of "large enough", but study of
a number of examples shows that large enough means a complete
set of commuting operators, and an equal number of operators
conjugate to them \cite{review-abhay,abhay-leshouches,tate-time}.
Now, as in ordinary quantum field theory, it
is very unlikely that any two operators defined at different physical
times commute.  Thus, we may postulate that the largest set of
operators for which reality conditions can be imposed for the physical
theory are two conjugate complete sets defined at a single moment
of physical time.

Thus, we could define an inner product by imposing the reality
conditions
for a complete set of operators
$\hat {\cal O}[\{ \hat A,\hat E,\hat \phi\}, \tau  ]_{g.f.}$
at any physical time $\tau$.  Now, the nicest situation would be if the
resulting inner products were actually independent of $\tau$.
However,
there are reasons, some of which are discussed  below,
to believe that this may not  be realized in full
quantum gravity.  If this is the case then the most natural
assumption
to make is that the physical inner product must be determined by a
complete set of operators at the initial time, $\tau=0$, as that will
correspond to the time of preparation of the physical system.

However, by (44) we see that to impose the reality conditions
on the operators
$\hat {\cal O}[\{ \hat A,\hat E,\hat \phi\}, \tau  ]_{g.f.}$ at
$\tau=0$ is to simply impose the diffeomorphism invariant
reality conditions for the non-clock fields.  Thus, we may
propose that for any two physical states $\Psi$ and $\Psi^\prime$
in ${\cal H}_{g.f.}$
\f
< \Psi | \Psi^\prime >_{g.f.} =
<\Theta \circ\Psi |\Theta \circ \Psi^\prime >_{diffeo}
\ff
As a result, we  may conclude that the physical expectation
value of the
operator that measures the diffeomorphism invariant
quantity ${\cal O}[\{ A,E,\phi \} ]_{diffeo}$ at the time $\tau$ in the
state $\Psi [\{ \gamma , \phi \} , \tau]$ in ${\cal H}_{g.f.}$
is given by
\f
<\Psi |\hat{\cal O}
[\{ \hat A,\hat E,\hat \phi\}, \tau  ]_{g.f.} |\Psi >_{g.f.}=
<\Theta \circ \Psi | \Theta \circ
\left ( \hat{\cal O}[\{ \hat A,\hat E,\hat \phi\}, \tau  ]_{g.f.}
|\Psi >\right )_{ diffeo}
\ff

\subsection{A word about unitarily}

The reader may notice that in fixing the physical inner product
there was one condition I might have imposed, but did not.
This was that the operator $\hat{\cal W}$ that generates evolution
for the non-clock fields be hermitian.  The reason for
this is one aspect of the
conflict between the notions of time in quantum theory and general
relativity.  From the point of view of quantum theory, it is natural
to assume that the time evolution operator is unitary.  However, this
means that all  physical states of the form
$\Psi[\{ \gamma , \phi \} , \tau ]$ exist for all physical
clock times  $\tau$.  This directly contradicts the situation in classical
general relativity, in which for every set of
initial data which solves the constraints
for compact $\Sigma$ (and which satisfies the positive energy
conditions) there is a time $\tau$ after
which the spacetime has collapsed to a final singularity so that
no physical
observable could be well defined.

Furthermore, we may note that we may not be free to choose the
physical inner product such that $\hat{\cal W}$ is hermitian.  The
physical inner product has already been restricted by the reality
conditions applied to a certain set of observables of the
diffeomorphism invariant theory.  As $\hat{\cal W}$  is
to be defined through a limit of a regularization procedure, it is
probably best to fix the inner product first and then define
the limit inside this inner
product space\footnote{Note that this is different from the problem
of finding the kernel of the constraints, for which it is
sufficient to define the limit in a pointwise topology,
as was done in \cite{tedlee,carlolee}. There we were content
to let the limit be undefined on the part of the state space not
in the kernel. As we now want to construct
the whole operator  we probably need
an inner product to control the limit.}.  However, then
it is not obvious that the
condition that $\hat{\cal W}$ be hermitian will be consistent
with the  conditions that determine the inner product.

How a particular formulation of quantum  gravity
resolves these conflicts is a dynamical problem.   This is, indeed,
proper, as the evolution operator for quantum gravity could be
unitary only if the quantum dynamics avoided
complete gravitational
collapse in every circumstance, and whether this is the case or not is
a
dynamical problem.  The implications of this situation will be the
subject of section 6.

\subsection{The physical quantum theory without gauge fixing}

As in the classical theory, once we have the gauge fixed theory it is
easier to see how to construct the theory without gauge fixing.
Here I will give no technical details, but only sketch the
steps of the construction.

In the Dirac approach, we first construct the kinematical
state space, which consists of all states with
a general dependence on the variables, of the form
$\Psi [ \gamma , \phi , T(x)] _{Dirac}$.  Instead of (33) and (34) we
have the defining relations
\f
\hat{T}(x) \Psi  [ \gamma , \phi , T(x)] _{Dirac}=
T(x) \Psi  [ \gamma , \phi , T(x)] _{Dirac}
\ff
and
\f
\hat{\cal E} (x) \Psi  [ \gamma , \phi , T(x)] _{Dirac}=
-\imath \hbar {\delta \over  \delta T(x) }
\Psi [ \gamma , \phi , T(x)] _{Dirac}.
\ff
The physical state space, ${\cal H}_{Dirac}$ then consists
of the subspace of states that satisfy the full set of constraints,
\f
\hat{\cal D}(v)\Psi_{Dirac} =0
\ff
and
\f
\lim_{\epsilon \rightarrow 0} \hat{\cal C}^\epsilon (N)
\Psi_{Dirac} =0
\ff
for all $v^a$ and $N$.  As in the gauge fixed formalism, we will be
interested only in those solutions that arise from initial data
of the non-clock fields, so that they can represent states prepared at
an initial clock time.  Thus, we will be interested in the subspace
of Dirac states such that
\f
\Psi [\{ \gamma , \phi , T (x)=0\} ]_{Dirac}=
\Psi [\{ \gamma , \phi \} ]_{diffeo}  .
\ff
is normalizable in an appropriate inner product that gives a
probability measure to the possible preparations of the system
we can make at the initial time.

There is a complication that arises in the case of the full
constraints, because (51) is a second order equation in
$\delta / \delta T(x)$.  This is the familiar problem of the doubling
of solutions arising from the Klein-Gordon like form of the full
Hamiltonian constraint.  In the gauge fixed quantization we studied
in section 3.2-3.4 this problem did not arise because the reduced
Hamiltonian constraint was first order in the derivatives of the
reduced time variable $\tau$.

However, there is a way to deal with this problem in the full, Dirac,
quantization, because we have the constant of motion ${\cal E}$
defined by (17).  We can use this to impose a positive frequency
condition on the physical states.
Thus, using $\hat{\cal E}$ we can
split the Hilbert space
${\cal H}_{Dirac}$,  to be
the direct product of two subspaces,  ${\cal H}_{Dirac}^\pm$,
where ${\cal H}_{Dirac}^+$ is spanned by the
eigenstates of $\hat{\cal E}$ whose
eigenvalues have positive real part, and ${\cal H}_{Dirac}^-$ is
spanned by the eigenstates
of $\hat{\cal E}$ with
 eigenvalues with negative real part.
Associated to this splitting we have projection
operators $P^\pm$ that project onto
each of these subspaces\footnote{Note that I have not assumed
that $\hat{\cal E}$ is hermitian, for the reasons discussed
in the previous section.}.   From now on, we will restrict
attention to states and operators in the positive frequency
part of the physical Hilbert space.

As in the gauge fixed case we thus define  a map
\newline
$\Theta : {\cal H}_{ Dirac}^+ \rightarrow {\cal H}_{diffeo}$
by
\f
\left (\Theta \circ \Psi_{Dirac} \right ) [\{ \gamma , \phi \} ]
=\Psi [\{ \gamma , \phi , T (x)=0\} ]_{Dirac}
\ff
Further, if there is a unique solution to the constraints that satisfies
also the positive frequency condition,
\f
P^+\Psi_{Dirac}=\Psi_{Dirac}
\ff
 there is a corresponding inverse map
$\Lambda : {\cal H}_{diffeo}\rightarrow {\cal H}^+_{Dirac} $
which takes each state in ${\cal H}_{diffeo}$ into its positive
frequency evolution under the full set of constraints.

The operators on this space, which we can call the Dirac operators are
as well solutions to the full set of constraints,
\f
\left [ \hat{\cal D}  (v), \hat{\cal O}_{Dirac} \right ] =0
\ff
\f
\lim_{\epsilon \rightarrow 0}
\left [ \hat{\cal C}^\epsilon  (N), \hat{\cal O}_{Dirac} \right ] =0
\ff
We will impose as well the positive frequency condition,
\f
\left [ P^+, \hat{\cal O}_{Dirac} \right ] =0
\ff
which
converts (56) from a second order to a first order
functional
differential equation.

We may then seek to use these equations to extend diffeomorphism
invariant operators on ${\cal H}_{diffeo}$, which act only on
the non-clock degrees of freedom, to positive frequency Dirac
operators.  That is, given an operator
$\hat{\cal O}
[\{ \hat E ,\hat A, \hat \phi \} ]_{diffeo}$
we seek operators of the form
$\hat{\cal O}
[\{ \hat E ,\hat A, \hat \phi , T(x)  ,\hat {\cal E} ]\}] _{Dirac}$
which solve (55), (56) and (57) which have the property that
\f
\hat{\cal O}
[\{ \hat E ,\hat A, \hat \phi , T(x) =0 ,\hat {\cal E} ]\} ]_{Dirac}
=\hat{\cal O}
[\{ \hat E ,\hat A, \hat \phi \} ]_{diffeo}
\ff

Furthermore, we can construct quantum analogues of the
evolving constants of motion (32).  These are corresponding one
parameter families of Dirac observables
$\hat{\cal O}^\prime[
\{ \hat E ,\hat A, \hat \phi , T(x) =0 ,\hat {\cal E} ]\} ](\tau )_{Dirac}$
that satisfy (55), (56) and (57)
(with $\tau$,
again, treated just as a parameter) and the condition
\f
\hat{\cal O}^\prime [\{ \hat E ,\hat A, \hat \phi , T(x)=0 ,
\hat {\cal E} ]\} ](\tau )_{Dirac} =
 \hat{\cal O}
[\{ \hat E ,\hat A, \hat \phi , T(x) =\tau ,\hat {\cal E} ]\} ]_{Dirac}
\ff

As in the classical case, one can relate these operators also to
the operators of the gauge fixed theory.  However, as there
are potential operator ordering problems that come from the
operator versions of the substitutions in (32), and as the gauge
fixed and Dirac operators act on different state spaces, it is
more convenient to make the definition in this way.

We may note that these may  not be all of the physical operators
of the Dirac theory, as there may be operators that satisfy the
constraints and positive frequency condition for which there
is no diffeomorphism invariant operator of only the non-clock
fields such that (58) holds.  But this is a large enough set
 to give the theory a physical
interpretation based on the use of the clock fields.

To finish the construction of the Dirac formulation,
we must give the physical inner product.  The same argument
that we gave in the gauge fixed case leads to the conclusion that we
may impose a physical inner product $<|>_{Dirac}$ such that,
if $\Psi$ and $\Phi$ are two
elements of ${\cal H}^+_{Dirac}$
\f
<\Psi|\Phi>_{Dirac} =
<\Theta \circ \Psi |\Theta \circ \Phi>_{diffeo}
\ff
For the reason just stated, this may not determine the whole
inner product on ${\cal H}_{Dirac}^+$, but it is enough to do
some physics because we may conclude that
in the Dirac formalism the expectation value of the operator
that corresponds to measuring the diffeomorphism
invariant quantity
${\cal O}_{diffeo}$ of the nonclock field a physical
time $\tau$ after the preparation of the system in the state $|\Psi >$
(which, by definition is in ${\cal H}^+_{ Dirac}$)  is
\f
<\Psi |\hat{\cal O}^\prime(\tau )_{Dirac}|\Psi>_{Dirac}=
<\Theta \circ \Psi|\Theta \circ
\left ( \hat{\cal O}^\prime (\tau )_{Dirac}|\Psi> \right )_{diffeo}.
\ff

\section{Outline of a measurement theory for quantum cosmology}

I will now move away from technical problems, and consider the
question of how a theory constructed according to the lines
of the last two sections could be interpreted physically.
In order to give an interpretation of a quantum theory it is necessary
to
describe what mathematical operations in the theory correspond to
preparation of the system and what mathematical operations
correspond
to measurement.  This is the main task that I hope to fulfil here.
I should note that I will phrase my discussion entirely in the
traditional language introduced by Bohr and Heisenberg concerning
the interpretation of quantum mechanics.  As we will see, with
the appropriate modifications, there is no barrier to using this
language in the context of quantum cosmology.
However, if the reader prefers
a different language to discuss the interpretation of quantum
mechanics,
whether it be the many worlds interpretation or a statistical
interpretation, she will, as in the case of ordinary quantum
mechanics,
be able to rephrase the language appropriately.

The  interpretation of quantum cosmology that I would like to
describe
is based on
the following four principles:

{\bf A)  The measurement theory must be completely
spacetime diffeomorphism
invariant.}    The interpretation must respect the
spacetime diffeomorphism
invariance of the quantum theory of gravity.  Thus, we must build
the interpretation entirely on physical states and physical
operators.

{\bf B) The reference system, by means of which we locate where
and when in the universe measurements take place, must be
a dynamical component of the quantum matter plus
gravity system on
which our quantum cosmology is based. }   This is a consequence
of the first principle, because the diffeomorphism invariance
precludes the meaningful use of any coordinate system that
does not come from the configuration of a dynamical variable.

{\bf   C) As we are are studying a quantum field theory, any
measurement
we can make on the system must be a partial measurement.}   This is
an important point whose implications will play a key role in what
follows.  The argument for it is simple: a quantum field theory has
an infinite number of degrees of freedom.  Any measurement that
we make returns a finite list of numbers.  The result is that any
measurement made on a quantum field theory can only result in
a partial determination of the state of
the system.

{\bf D)  The inner product is to be determined by requiring that a
complete
set of physical observables for the gravity and matter
degrees of freedom satisfy the reality
conditions at the initial physical time corresponding to preparation of
the state.  }

For concreteness I will phrase the measurement theory in terms of
the particular type of reference frames and clock fields described
in the last two sections.  However, I will use a language that
can refer to either the gauge fixed formalism described in
subsections 3.2-3.4 or the Dirac formalism described in
subsection 3.6.  I will use a general subscript $phys$ to refer
to the physical states, operators and inner products of either
formalism.  If one wants to specify the gauge fixed formalism
then read $phys$ to mean $g.f.$ so that operators
$\hat{\cal O}(\tau )_{phys.}$ will mean the gauge fixed
operators $\hat{\cal O}[\hat{E},\hat{A},\hat{\phi},
\hat {\cal E},  \tau ]_{g.f.}$ defined in subsection 3.3.
Alternatively, if one wants to think in terms of the Dirac formalism
then read $phys$ to mean $Dirac$ everywhere, so that the
operators $\hat{\cal O}(\tau )_{phys.}$ refer to the $\tau$
dependent "evolving constants of motion"
$\hat{\cal O}^\prime (\tau)_{Dirac}$.  Furthermore,
I will always assume that reference is being made to states
and operators in the positive frequency subspace of the
Dirac subspace.  I will use this notation as
well in section 6.

Thus, putting together the results of the last two sections, I shall
assume, for purposes of illustration,
that we have available at least two sets of $\tau$ dependent
physical observables $\hat{\cal A}^I (\tau )_{phys}$ and
 $\hat{ T}^I (\tau )_{phys}$ which measure, respectively,
the areas of the simplices of the reference frame, and parallel
transport around them, at a physical time $\tau$.

However, while I refer to a particular
form of clock dynamics and a particular set of observables, I
expect that the interpretation given here can be applied to
any theory in which the physical states, observables and inner
product are related to their diffeomorphism invariant  counterparts
in the way described in the last section.

Let us now begin with the process of preparing a system for an
observation.

\subsection{Preparation in quantum cosmology}

Let me assume that at time $\tau=0$ we make a preparation
prior to performing some series of measurements on
the quantum gravitational field.  This means that we
put the quantum fields
which
describe the temporal and spatial reference system into appropriate
configurations so that the results of the measurements will be
meaningful.  There are two parts to the preparation: arrangement
of the spatial reference system and synchronization of the
physical clocks.

As in ordinary quantum mechanics, we can assume that we, as
observers,
can move matter around as we choose in order to do this.  This
certainly
does not contradict the assumption that the whole universe
including
ourselves could be described by the quantum state $\Psi$ for, if it
did,
we would be simply unable to do quantum cosmology because,
{\it ipso facto} we are in the universe and we do move things such
as clocks and measuring instruments around more or less as we
please.

In ordinary quantum mechanics the act of preparation may be
described by projecting the quantum states of the reference system
and
measuring instruments into appropriate states, after which the
direct
product with the system state is taken.
In the case of a diffeomorphism
invariant theory we cannot do this because there is no basis
of the diffeomorphism invariant space ${\cal H}_{diffeo}$ whose
elements can
be written
as  direct products of matter states and gravity states.  Thus,  the
requirement of diffeomorphism
invariance has entangled the various
components
of the whole system even before any interactions occur.

However, this entanglement does not prevent us from describing in
quantum mechanical terms the preparation of the reference system
and clock fields.  What we must do is describe the preparation by
projecting the physical states into appropriate subspaces,
every state of which describes a physical situation in which
the matter and clock fields have been prepared appropriately.

Let us begin with synchronizing the clocks.
We may assume that we are able to synchronize a
field
of clocks over as large a volume of the universe as we please, or
even over
the whole universe (if it is compact).  The difficulty of doing this is,
{\it a priori} a practical problem, not a problem of principle.   So we
assume that
we may synchronize our clock field so that there is a spacelike
surface everywhere on which  $T(x)=0$.

In terms of the formalism, this act of preparing the clocks corresponds
to assuming that the state is normalizable in the inner product
defined by either (46) or (60).  That is, there may be states in
either the solution space to the gauge fixed constraints
or the Dirac constraints that are not
normalizable in the respective inner products defined by
the maps to the diffeomorphism invariant states of the
non-clock fields.  Such states cannot correspond to
preparations of the matter and gravitational fields made
at some initial time of the clock fields.

Once the clocks are synchronized we can prepare the spatial
reference frame.  As described in subsections 2.2 and 2.3
we do this by specifying that
the $N$ surfaces are arranged as the faces of a simplicial complex
$\cal T$.  In the
formalism this is described by the statement that the state of the
system
is  to be further
restricted to be in a subspace ${\cal H}_{ phys , \cal T}$
of ${\cal H}_{phys}$.

It is interesting to note that, at least in principle, the
preparation of the spatial and temporal reference frames
can be described
without making any assumption about the quantum state of the
gravitational field.  Of course, this represents an ideal case, and in
practice  preparation for a measurement in quantum
cosmology will usually involve fixing some degrees of freedom
of the gravitational field.  But, it is important to note that this is
not required in principle.  In particular, no assumption need
be made restricting the gravitational field to be initially in anything
like a classical or semiclassical configuration to make the
measurement
process meaningful.

This completes the preparation of the spatial and
temporal reference frames\footnote{Note that,
in a von Neumann type description
of
the measuring process, we must also include the measuring
instruments
in the description of the system.  These are contained in the
dependence
of  the physical states on additional
matter fields in the set $\phi$ that represent the actual
measuring instruments.
I will not here go through the details of adopting the
von Neumann description of measurement to the present case, but
there is certainly no obstacle to doing so.   However, as
pointed out by Anderson \cite{arley} it is necessary
to take into account the fact that a real measuring interaction
takes a finite amount of physical time.  There is also no obstacle
to including in the preparation
as much information about our own existence as may be
desired, for example, there is a
subspace of ${\cal H}_{phys, {\cal T}}$ in
which we are alive, awake,
all our measurement instruments are prepared  and we are
in a mood to do an experiment.
There is no problem with assuming this
and requiring that the system be initially in this subspace.
However, as in ordinary quantum mechanics, as long as we do not
explicitly make any measurements on ourselves, there is no
reason to do this.}.

\subsection{Measurement in quantum cosmology}

After preparing the system we may want to wait a certain
physical time
$\tau$
 before making a measurement.
Let us suppose, for example, that we want to measure the
area of one of the surfaces picked out by the
spatial reference system at the time $\tau$ after
the preparation.    How
are we to describe this?   To answer this question we need
to make a postulate, analogous to the usual
postulates that connect measurements to the actions
of operators in quantum mechanics.    It seems most
natural to postulate the following:
\blankline
\noindent
{\bf Measurement postulate of quantum cosmology}:
 {\it The
operator that corresponds to the making
of a measurement of the spatially diffeomorphism invariant
quantity represented by $\hat{O}_{diffeo} \in {\cal H}_{diffeo}$
at the timhat the clock field reads $\tau$, is the
physical
(meaning either gauge fixed or Dirac)
operator $\hat{O}_{phys} (\tau)$.
Thus, we postulate that the expected value
of
making a measurement of the quantity $\cal O$ at a time $\tau$ after
the preparation in the physical state $|\Psi >$ is given by}
$ <\Psi | \hat{O}_{phys} (\tau) |\Psi >_{phys}$.

It is consistent with this to postulate also that: {\it the only
possible values which may result from a measurement of the
physical quantity $\hat{O}_{phys} (t)$ are its
eigenvalues, which may be found  by
solving the physical
states equation}
\f
\hat{O}_{phys} (\tau)|\lambda (\tau ) >_{phys} = \lambda
(\tau)
|\lambda (\tau) >_{phys}
\ff
inside the physical state space ${\cal H}_{phys}$.

Having described how observed quantities correspond
to the mathematical expressions of quantum cosmology, we have
one more task to fulfil to complete the description of the measurement
theory.  This is
to confront
the most controversial part of measurement theory, which is
the question of
what happens to the quantum state after we make the measurement.
In ordinary quantum mechanics there are two points of view about
this,
depending on whether one wants to employ the projection
postulate or some version of the relative state idea of
Everett.  This choice is usually, but
perhaps not necessarily, tied to the philosophical point
of view that one holds
about the quantum state.  If one believes, with Bohr and von
Neumann,
that the quantum state is nothing physically real, but only
represents our information about the system, then there is
no problem with speaking in terms of the projection postulate.
There is, in this way of speaking, only an abrupt change in the
information that we have about the system.  Nothing physical
changes, i.e. collapse of the wave function is not a physical
event or process.

On the other hand, if one wants to take a different point of view
and postulate that the quantum state is directly associated with
something real in nature, the projection postulate brings with
it the well known difficulties such as the
question of in whose reference frame
the collapse takes place. There are then two possible
points of view that may be taken. Some authors, such as
Penrose \cite{Roger-book},
take this
as a physical problem, to be solved by a theory that is to
replace, and explain, quantum mechanics.  Therefore,
these authors want
to accept the collapse as being something that physically
happens.  The other point of view is to keep the postulate
that the quantum state is physically real but to give an
interpretation of the theory that does not involve
the projection postulate.
In this case one has to describe measurement in terms of the
correlations that are set up during the measurement process
between the quantum state of the measuring instrument
and the quantum state of the system as a result
of their interaction during the measurement.

Of course, these two hypotheses lead, in principle, to different
theories as in the second it is possible to imagine doing experiments
that involve superpositions of states of the observer, while in the
first case this is not possible.  Nevertheless, there is a large
set of cases in which the predictions of the two coincide.  Roughly,
these are the cases in which the quantum state, treated from the
second point of view, would decohere.  Indeed, if one takes
the second point of view then some version of decoherence is
necessary to recover what is postulated from the first point
of view, which is that the observer sees a definite outcome to
each experiment.

I do not intend to settle here the problem of which of these
points of view  corresponds most closely to nature.  However, I
would like to make two claims which I believe, to some extent,
diffuse the conflict.  First,  I would
like to claim that whichever point of view one takes, something
like the statement of the projection postulate plays a role.
Whether it appears as a fundamental statement of the interpretation,
or
as an approximate and contingent statement
which emerges only in the case of decoherent states or histories,
the connection to what real observers see can be described
in terms of the projection postulate, or something very much
like it.  Second, I would
like to claim  that the situation is not different in quantum
cosmology then it is in ordinary quantum mechanics.  One
can make either choice and, in each case, something like
the projection postulate must enter when you discuss the
results of real observations (at least as long as one is not
making quantum observations on the brain of the observer.)

With these preliminaries aside, I will now state how
the projection postulate can be phrased so that it
applies in the quantum cosmological case:

\blankline
\noindent
{\bf Cosmological
projection postulate:}
{\it Let $\hat{\cal O}_I (\tau_0)_{phys}$ be a finite
set of physical operators that
mutually commute and hence correspond to a set of measurements
that can be made simultaneously at the time $\tau_0$.  Let us assume
that the reference system has been prepared so that the
system before the measurements is in the subspace
${\cal H}_{phys , {\cal T}}$.  The results
of the observations will  be a set of eigenvalues
$\lambda_I (\tau_0)$.  For
the
purposes of making any further measurements,
which would correspond
to values of the physical clock field $\tau$ for
which $\tau > \tau_0$, the
quantum
state can be assumed to be projected at the physical clock time
$\tau_0$ into the subspace
${\cal H}_{phys, {\cal T}, \lambda (\tau_0)}
\subset {\cal H}_{phys, {\cal T}}$ which
is spanned by all the eigenvectors of the operators
$\hat{\cal O} _I(\tau _0)_{phys}$
which correspond to the
eigenvectors
$\lambda_I(t_0)$.}

That is, one is to project the state into the subspace at the time of
the measurement, and then continue with the evolution defined
by the Hamiltonian constraint.

\subsection{ Discussion}

I would now like to discuss three objections that might
be raised concerning the application of the projection postulate to
quantum cosmology.  Again, let me stress that my goal here is not
to argue that one must take Bohr's point of view over that of
the other interpretations.  I only want to establish that
if one is happy with Bohr in ordinary quantum mechanics one
can continue to use his point of view in quantum cosmology.  The
only strong claim I want to make is that the statement
sometimes made, that
one is required to give up Bohr's point of view when one comes to
quantum cosmology, is false.

{\bf First objection:} {\it Bohr explicitly states that
the measurement apparatus must be described classically, which
requires that it be outside of the quantum system being studied.}
I believe that this represents a misunderstanding of Bohr which,
possibly, comes from combining what Bohr did write with an
assumption
that he did not make, which is that the quantum state is
in one to one correspondence with something physically real.
For Bohr to have taken such a  realistic point of view about the
quantum state would have been to directly
contradict his fundamental point of view about physics,
which is that it does not involve any claim to a realistic
correspondence
between nature and either the mathematics or the words we use to
represent the results of observations we make.  Instead, for Bohr,
physics is an extension of ordinary language by means of which
we describe to each other the results of certain activities we do.
Bohr takes it as given that we must use classical language to
describe the results of our observations because that is
what real experimentalists do.  Perhaps the weakest point of
Bohr is his claim that it is necessary that we do this, but even if
we leave aside his attempts to establish that, we are still left with
the fact that up to do this day the only language that we actually
do use to communicate with each other what happens when we
do experimental physics uses certain classical terms.

Furthermore, rather than insisting that the measuring instrument is
outside of the quantum system, Bohr insists repeatedly that the
measuring instrument is an inseparable part of the entire system
that is described in quantum mechanical terms.  He insists that
we cannot separate the description of the atomic system from a
description of the whole experimental situation, including
both the atoms and the apparatus.

Many people do not like this way of talking about physics.
My only point here is that there is nothing in this way of speaking
that prevents us from doing quantum cosmology.  After all, we
are in the universe, we are ourselves made of atoms, and we do
make observations and describe their results to each other in classical
terms.  That all these things are true are no more and no less
mysterious
whether the quantum state is a description of our observations made
of the spin of an atom or of the fluctuations in the
cosmic black body radiation.

{\bf Second objection:} {\it It is inconsistent with the idea
that
the quantum state describes the whole universe, including us,
to postulate that the result of a measurement  that we
make is one of the
eigenvalues
of the measured observable, because that is to employ a classical
description, while the whole universe is described by a quantum
state.}  To say so is, in my opinion, again to
misunderstand
Bohr and von Neumann and, again, to attempt to combine their
way of speaking about physics with some postulate about the
reality of the quantum state.  To postulate that the result of a
measurement
is an eigenvalue is to assume that the results of measurements may
be
{\it described} using quantities from the language of classical
physics.
The theory does not, and need not, explain to us why that is the case;
that we get definite values for the results of experiments
we do is taken as a primitive fact upon which we base
the interpretation of the theory.

Furthermore, to
assume that the results of measurements are {\it described}
in terms of the language of classical physics is not at all the same
thing as to make the (obviously false)
claim that the dynamics that governs the physics of
either the measuring instruments or ourselves is classical.

{\bf Third objection:} {\it The fact that we are in the universe might
lead to some problem in quantum cosmology
because a measurement of a quantum state would involve
a measurement of our own state. } There are two replies to this.

First, there is an interpretation of quantum mechanics, suggested by
von Neumann\cite{vonN}
and developed to its logical conclusion by Wigner\cite{wigner},
that
says that all we ever actually do is make observations on our own
state.
Wigner claims that there is something special about consciousness
which is
that we can experience only definite things, and not superpositions.
This is then taken to be
the explanation for why we observe the results of
experiments
to give definite values.  I do not personally believe this point of
view, but
the fact that it is a logically possible interpretation of quantum
mechanics
means that there can  be no logical problem with including ourselves
(and
all our measuring instruments and cats, if not friends) in
the description of the quantum state.

Second, we can avoid this problem, at least temporarily, if we
acknowledge
that all measurements we make in quantum cosmology are
incomplete
measurements.  In reality we never determine very much information
about the quantum state of the universe when we make a
measurement,
however we interpret it.  We certainly learn very little about our
own
state when we make a  measurement of the gravitational
field of the sort I
described
in section 2.

Thus, as long as we refrain
from actually describing experiments in which we make
measurements
on our own brains, we need not commit ourselves to any claims
about
the results of making observations on ourselves.  Again, the
situation here
is exactly the same  in quantum cosmology as it is in ordinary
quantum
mechanics-no worse and no better.
If there is a possible problem with making observations on
ourselves in quantum cosmology, it must occur in ordinary quantum
mechanics.  And it must be faced there as well, as it cannot matter
for
the resolution of such a problem whether, besides our brain,
Andromeda
or the Virgo cluster is also described by the quantum state.

Let me close this section with one comment.  Given the measurement
postulate above, and the results of the last two sections, we can
conclude that in fact the areas of surfaces are quantized in quantum
gravity.  For, without integrating the evolution equations,
we know that
$\hat{\cal A}^I(\tau=0)_{phys}=\hat{\cal A}^I(\tau=0)_{diffeo}$
and the latter operator, from the results of
\cite{carloobserves,antisymm} has a discrete spectrum.  Thus,
quantum gravity makes a physical prediction.  Note, further
that this result is independent of the form of the Hamiltonian
constraint and hence of the dynamics and the matter content of
the theory.

\section{The recovery of conventional quantum field theory}

The measurement theory given in the previous section
has not required any notion of classical or semiclassical states.  One
need only assume that it is possible to prepare the fields that
describe the spatial and temporal quantum
reference frame appropriately so that subsequent
measurements are meaningful.  One does not need to assume that
the gravitational field is in any particular state to do this.  Of
course,
there may be preparations that require some restriction on the state
of the gravitational degrees of freedom, but such an assumption
is not required in principle.   Further, the examples
discussed in the last sections show that there
are
some kinds of physical reference frames whose preparation requires
absolutely no restrictions on the gravitational field.

Having said this, we may investigate what happens to the
dynamics and the
measurement
theory if we add the condition that the state is semiclassical
in the gravitational degrees of freedom.   I will
show in this  section  that by making the assumption
that the
gravitational field is in a semiclassical state we can recover quantum
field theory for the matter fields on a fixed spacetime background.
Thus, quantum cosmology, whose
dynamics is contained in the quantum constraint equations, and
whose
interpretation was described in the previous section, does have a
limit
which reproduces conventional quantum field theory.

Here I will only sketch a version of the demonstration, as my main
motive here is to bring out an interesting point regarding a
possible role for the zero point energy in the transition from
quantum gravity to ordinary quantum field theory.   For
simplicity,
I will also drop in this section the assumption that the state is
diffeomorphism
invariant, as this will allow me to make use of published results
about
the semiclassical limit of quantum gravity in the loop
representation.
However, I will continue to treat the Hamiltonian constraint in the
gauge fixed formalism of sections 3.2-3.4

I will make use of the results described in \cite{weaves} in which it
was
shown
that, given a fixed three metric $q^{ab}_0$, whose curvatures are
small
in Planck units, we can construct, in
the loop representation,  a nonperturbative
quantum state of the gravitational field
that approximates
that classical metric up to terms that are small in Planck units.

Such a state can be described as follows\footnote{More details
of this construction are given in \cite{weaves}.
See also \cite{abhay-leshouches,review-carlo,review-lee}.}.
Given the volume element
$\sqrt{q_0}$, let me distribute points randomly on $\Sigma$ with a
density $1/l_{Planck}^3 \times (2/\pi )^{3/2}$.
Let me draw a circle around each point
with a radius given by $\sqrt{\pi \over 2} l_{Planck}$
with an orientation in space that
is random, given the metric $q^{ab}_0$.  As the curvature is
negligible over each of these circles, this is well defined.  Let me
call the collection of these circles $\Delta= \{ \Delta_I\} $,
where $I$ labels them.

Let me then define the {\it weave state associated to} $q^{ab}_0$ as
the {\it characteristic state } of the set of loops $\Delta$.  This
is denoted $\chi_\Delta$ and, for a non-selfintersecting loop
$\Delta$,  it is defined so that $\chi_\Delta [\alpha ]$
is an eigenstate of the area operator ${\cal A}[{ \cal S}]$ which
measures
the area of the arbitrary surface with eigenvalue $l_{Planck}^2/2$
times
the number of times the loop $\Delta$ intersects the surface $\cal
S$.
The result is that $\chi_\Delta [\alpha ]$ is equal to one if $\alpha$
is
equivalent to $\Delta$ under the usual rules of equivalence of loops
in the loop representation and is equal to zero for most other
loops
$\alpha$ (including all other distinct non self-intersecting loops.)

I will also assume that the weave is chosen to have no intersections,
in which case
\f
\lim_{\epsilon \rightarrow 0} {\cal C}_{grav.}^\epsilon (x)
\chi_\Delta [\alpha ] =0
\ff
This will simplify our discussion.

I would now like to make the ansatz that the state is of the form
\f
\Psi [ \gamma ,\phi ,\tau]=
\chi_\Delta [\gamma  ] \Phi [\gamma ,\phi ,\tau] .
\ff
Note that, as I have dropped for the moment the requirement of
diffeomorphism invariance, I have also dropped the dependence
on the spatial reference frame field ${\cal S}$.

Now, let me assume, as an example, that the matter consists of
one scalar field, called $\phi (x)$,
with conjugate momenta
$\pi (x)$, whose contribution to the regulated Hamiltonian
constraint
is,
\begin{eqnarray}
\hat{\cal C}_\phi^\epsilon (x) &=& -
{1 \over 2} \int d^3y \int d^3z f^\epsilon(x,y) f^\epsilon(x,z)
\nonumber \\
&&\times \left [   \pi (y) \pi (z) +
\hat{T}^{ab}(y,z) \partial_a \phi(y)\partial_b \phi(z) \right ]  .
\end{eqnarray}

Let us focus on the spatial derivative term.   Let us assume that
the dependence of $\Phi [\gamma , \phi ,\tau]$ on the gravitational
field loops $\gamma$ can be neglected.  (This is exactly to neglect
the back reaction and the coupling of gravitons to the matter field.)
Let me assume also
that the support of the state on configurations on which the
scalar field is not slowly varying on the Planck scale (relative to
$q^{ab}_0$)
may be neglected.
Then it is not hard to show, following the
methods of \cite{weaves,review-lee} that as long as
the scalar fields are slowly varying on the Planck scale,
\begin{eqnarray}
&&\int d^3y \int d^3z f^\epsilon(x,y) f^\epsilon(x,z)
\hat{T}^{ab}(y,z) \partial_a \phi(y)\partial_b \phi(z)
\Psi [ \gamma ,\phi ,\tau ] \nonumber \\
&=&
\left ( \int d^3y \int d^3z f^\epsilon(x,y) f^\epsilon(x,z)
\hat{T}^{ab}(y,z) \chi_\Delta [\gamma ] \right )
\partial_a \phi(y)\partial_b \phi(z)  \Phi [\gamma ,\phi ,\tau]
\nonumber \\
&=&
\sum_I \sum_J \int ds \int dt f^\epsilon (x, \Delta_I(s))
f^\epsilon (x, \Delta_J(t)) \dot{\Delta}^a_I(s) \dot{\Delta}^b_I(t)
\nonumber \\
&&\times
\left ( \sum_{\mbox{routings}}\chi_\Delta [\gamma \circ
\gamma_{x,y}] \right )
\partial_a \phi( \Delta_I(s))\partial_b \phi(\Delta_J(t))
\Phi [\gamma ,\phi ,\tau]    \nonumber \\
&=&
det(q_0)q^{ab}_0 (x) \partial_a \phi(x)\partial_b \phi(x)
\chi_\Delta [\gamma ] \Phi [\gamma ,\phi ,\tau]
+ O (l_{Planck}^2 \partial_a \phi )
\end{eqnarray}
Putting these results together, we have shown that if we
make an ansatz on a state of the form of (64) then, neglecting
the dependence of $\Phi$ on $\gamma$, the regulated Hamiltonian
constraint (36) is equivalent to
\f
\imath
{d \Phi [\gamma ,\phi ,\tau]  \over d\tau}
 =  \sqrt{2\mu} \int d^3 x  \sqrt{ \left [ {1\over 2} \hat{\pi}^2 (x) +
{1\over 2} det(q_0) q^{ab}_0 (x)
\partial_a \phi (x) \partial_b \phi (x) \right ]  }
\Phi[\gamma ,\phi ,\tau]   + ...
\ff

This does not yet look like the functional Schroedinger equation
for the scalar field.  However, we may recall that formally the
expression inside the square root is divergent.    However, it
may not be actually divergent because in computing (66) we have
assumed that the scalar field is slowly varying on the scale of
the weave.  If we investigate the action of (65) on states which
have support on $\phi (x)$ that are fluctuating on the Planck
scale, we can see that in the limit that the
regulator is removed the effect of the $T^{ab}$'s is to
insure that the the terms in $(\partial_a\phi )^2$ only act
at those points which are on  the lines of the weave.  That is, in
the limit of small distances we have a description of a scalar
field propagating on a one dimensional subspace of $\Sigma$
picked out by the weave.  That is, on scales much smaller than
the Planck scale the scalar field is propagating as a
$1+1$ dimensional scalar field.

The result must be to  cut off the divergence
in the zero point energy coming formally from the scalar field
Hamiltonian.  The effect of this must be the following:  If we
decompose the scalar field operators into creation and annihilation
operators defined with respect to the background metric
$q_0^{ab}$ that the weave corresponds to, then the
divergent term  in the zero point energy must cut off at
a scale of $M_{Planck}$.  As such, we will have, if we
restrict attention to the action of (67)
on states that are slowly varying on the
Planck scale
\begin{eqnarray}
{1\over 2} \hat{\pi}^2 (x) &+&
{1\over 2} det(q_0) q^{ab}_0 (x)
\partial_a \phi (x) \partial_b \phi (x)  = a M_{Planck}^4
\nonumber \\
 &&+ :     {1\over 2} \hat{\pi}^2 (x) +
{1\over 2} det(q_0) q^{ab}_0 (x)
\partial_a \phi (x) \partial_b \phi (x) :
\end{eqnarray}
where $:...:$ means normal ordered with respect to the
background metric and $a$ is an unknown constant that
depends on the short distance structure of the weave.

The reduced Hamiltonian constraint now becomes,
\begin{eqnarray}
\imath
{d \Phi [\gamma ,\phi ,\tau]  \over d\tau}
 &=&  \sqrt{2 a \mu} M_{Planck}^2
\nonumber \\
  &\times& \int d^3 x
\sqrt{  1
+ {1 \over a M_{Planck}^4}: \left (
{1\over 2} \hat{\pi}^2 (x) +
{1\over 2} det(q_0) q^{ab}_0 (x)
\partial_a \phi (x) \partial_b \phi (x) \right ) : }
\Phi[\gamma ,\phi ,\tau]   + ...
\nonumber \\
&=&   \sqrt{2 a \mu} M_{Planck}^2 V
\Phi[\gamma ,\phi ,\tau] \nonumber \\
 &+&
{\sqrt{\mu}\over M_{Planck}^2 \sqrt{2a}}
\int d^3x \sqrt{q_0}
: \left (  {1\over 2} \hat{\pi}^2 (x) +
{1\over 2} det(q_0) q^{ab}_0 (x)
\partial_a \phi (x) \partial_b \phi (x) \right ) :
\Phi[\gamma ,\phi ,\tau]  \nonumber \\
 &+& ...
\end{eqnarray}
where $V=\int \sqrt{q_0}$ is the volume of space.

Thus, only after taking into account
the very large zero point energy do
we recover conventional quantum theory for low energy
physics.

Before closing this section, I would like to make three comments on
this
result.

1)  Note that the theory we have recovered is Poincare invariant,
even
if the starting point is not!   We may note that
the weave state
$\chi_\Delta$
is {\it not} expected to be the vacuum state of quantum gravity
because
it is a state in which the spatial metric is sharply defined.  What we
need
to describe the vacuum is a Lorentz invariant state which which is
some kind of minimal uncertainty wave packet in which the three
metric
and its conjugate momenta are equally uncertain.  A state that has
these properties, at least at large wavelengths, can be constructed
by
dressing $\chi_\Delta$ with a Gaussian distribution of large loops
that
correspond to a Gaussian distribution of
virtual gravitons \cite{graviton}.
It is interesting to note that at the level when we neglect the
back-reaction
and the coupling to gravitons, the Poincare invariant matter quantum
field theory is nevertheless recovered by using the weave state
$\chi_\Delta$ as the background.
However, before incorporating quantum back-reaction and the coupling
to gravitons we must replace $\chi_\Delta$ in (64) with a good
approximation
to the vacuum state, such as is described in \cite{vacuum}.

2)  Can we add a mass term and self-interaction terms for the
scalar
field theory?  The answer is yes, but to do so we must modify the
weave
construction in order to add intersections.  The reason is that the
scalar mass and self interaction is described by the term
$\hat{q}V(\hat{\phi} )$, where $\hat{q}$ is the operator
corresponding
to the determinant of the metric.
Using results about the volume operator in  \cite{review-lee},
it is easy to see that if we modify
the weave construction in order to add intersections, then the effect
of
this term, after regularization, is to modify
$\int d^3x N(x) {\cal C}_{matter} \Psi $ by the addition of the term
\f
l_{Planck}^3 \sum_i a(i) N(x_i) V(\phi (x_i))\Phi
\ff
where the sum is over all intersections involving three or more
lines,
$x_i$ is the intersection point and $a(i)$ are dimensionless numbers
of
order one that characterize each intersection.  Assuming that there
are
on the order of $a(i)^{-1}$ intersection points per Planck volume,
measured with respect to the volume element $q_0$  (which
is consistent with the weave construction described above as that
is the approximate number of loops) we arrive at an addition to
(69) of the form of $N(x) q(x) V(\phi (x))\Phi $.

Note that once we add intersections that produce volume it is no
longer true that the gravitational part of the Hamiltonian constraint
is solved by $\chi_\Delta$.  This is because there is now a term
in the back-reaction of the quantum matter field on the metric
coming from the local potential energy of the scalar field.
This is telling us that we now cannot neglect the back-reaction of the
quantum fields on the background metric to construct solutions of
the Hamiltonian constraint.

3)  Finally, let me note that the measurement theory of the
semiclassical
state (64) is already defined because we have a measurement
for  the full nonperturbative theory.
We therefore do not have to supplement the derivation of the
equations of quantum mechanics from solutions to the quantum
constraint equations of quantum cosmology with the
{\it ab initio} postulation of the standard rules of interpretation of
quantum field theory.  This is always a suspicious procedure as
those
rules rely on the background metric that is only a property of a
particular state of the form of (64); we
cannot then choose inner
products
or other aspects of the interpretative machinery to fit a particular
state.

In this case, since we already have an inner product and a set of
rules of interpretation defined for the full theory, what needs
to be done is to
verify that the usual quantum field theory inner product is recovered
from the full physical inner product defined by (46) in
the case that the
state is of the form of (64) . We have seen in section 3 that
this will be
the case to the extent to which ${\cal W}$ defined by (37) is
hermitian.  We see that in the approximation that leads to (69),
the contribution to ${\cal W}$ from the matter fields is hermitian
a long as the diffeomorphism inner product implies that the
operators for $\phi$ and $\pi$ are also hermitian.   In this
case, then,
the usual inner product of quantum field theory must
be recovered.

\section{Singularities in quantum cosmology}

Having established a physical interpretation for quantum cosmology
and
shown that it leads to the recovery of conventional quantum field
theory in appropriate circumstances, we now have tools with which
to address what is perhaps the key problem that any quantum
theory
of quantum gravity must solve, which is what happens to black holes
and
singularities in the quantum theory.

The main question that must be answered is to what extent the
apparent
loss of information seen in the semiclassical description of black hole
evaporation survives, or is resolved, in the full quantum theory.  The
key
point that must be appreciated to investigate this problem from the
fully quantum mechanical point of view is that the problem of loss of
information, or of quantum coherence, is a problem about time
because the question cannot be asked without assuming that there is
a meaningful notion of time with respect to which we can say information
or coherence is being lost. If we take an operational approach to the
meaning of time in the full quantum theory, along the lines that have
been developed here, then the loss of information or coherence, if it
exists at the level of the full quantum theory, must show up
as a limitation
on the possibility of completely specifying the quantum state of the
system by measuring the physical observables,
${\cal O}(\tau )_{phys}$ for sufficiently late $\tau$.

What I would like
to do in this section is to describe how singularities, if
they occur in the full quantum theory, will show up in the
action of these physical time dependent observables.
The result will be the formulation of two
conjectures about how singularities may show up in the quantum
theory,
which I will call the {\it quantum singularity conjecture} and the {\it
quantum cosmic censorship conjecture}.

While I will not try to prove these conjectures here, I also will argue
that there is no evidence that they may not be true.  It is possible
that quantum effects completely eliminate the singularities of the
classical
theory as well as the consequent losses of information and
coherence in the
semiclassical theory.  But, it seems to me, it is at least
equally possible that
the quantum theory does not eliminate the singularities.  Rather,
given
the formulation of the theory along the lines described
 in this paper, the occurrence of
singularities and loss of information, as formulated in these
two conjectures,  seems to be  compatible with both
the dynamics and interpretational framework of quantum cosmology.

The loss of information and coherence in the semiclassical theory
implies
a breakdown in unitarity in any process that describes a black hole
forming
and completely evaporating.   From a naive point of view, this would
seem
to indicate a breakdown in one of the fundamental principles of
quantum
mechanics.  Hence many discussions about this problem seem to
assume
that if there is to be a good quantum theory of
gravity it must resolve
this problem in such a way that unitarity is restored in the full
quantum
theory.  However, the results of the last several sections show
that unitarity
is not one of the basic principles of quantum cosmology.  It is not
because
unitarity depends on a notion of time that apparently cannot be
realized
in either classical or quantum cosmology.

 To put it most simply, if the concept
of time is no longer absolute, but depends for its properties on
certain
contingent facts about the universe, principally the existence of
degrees of
freedom that behave as if there is a universal and absolute
Newtonian
notion of time, then the same must
be true for those structures and principles that,
in the usual formulation of quantum mechanics,
are tied to the absolute background time of the Schroedinger
equation.
Chief among these are the notions of unitarity evolution and
conservation
of probability.

As I have argued in detail
elsewhere\cite{spaceandtime,napoli}, conventional
quantum mechanics, no less than
Newtonian mechanics, relies for its interpretation on
the assumption of
an absolute background time.  When we speak of conservation of
probability, or unitary evolution in quantum mechanics, we do not
have to ask whether something might happen to the clocks that
measure time that could make difficulties for our
understanding of the operational meaning of these concepts.  A single
clock could break, but the $t$ in Schroedinger's equation refers to
no particular clock but instead to an absolute time that is presumed to
exist independently of the both the physical system described
the quantum state and of the physical properties of any particular
clock. However, neither in classical nor in quantum cosmology does there
seem to be available such an absolute notion of time.  In any case, if
it
exists we have not found it.  If we then proceed by using an
operational
notion of time as I have done here then, because that clock must, by
diffeomorphism invariance, be a dynamical part of the system under
study, we must confront the question: what happens to the notion of
time
and to all that depends on it if something happens to the clock whose
motion is taken as the operational basis of time.  Of course, in the theory,
as in real life, in most circumstances in which a clock may fail we can
imagine constructing a better one.  The problem we really have to face
as theorists is not the engineering problem of modeling the best
possible
clock.  The problem we have to face is what the implications are for
the
theory if there are physical effects that can render useless
any conceivable
clock.

Of course, in classical general relativity there is such an effect; it is
called gravitational collapse.     Thus, to put the point in the simplest
possible way, in quantum cosmology a breakdown of unitarity need
not indicate a breakdown in the theory.  Rather, it may only indicate
a breakdown in the physical conditions that make it possible to speak
meaningfully of unitary evolution.  This will be the case
if there are quantum states in
which some of the physical
clocks and some of the components of the physical
reference frame that make  observations in
quantum cosmology meaningful cease to exist
after certain physical times.  This will not prevent us from
describing
the further evolution of the system in terms of operators
whose meaning is tied to the physical clocks that happen to
survive the gravitational collapse.  But it will prevent us
from describing
that evolution in terms of the unitary evolution of the initial
quantum
state.

In this section I proceed in two steps.  First, before we describe
how singularities may show up in the operator algebra of the quantum
theory, we should see how they manifest themselves in terms of the
observable algebra of the classical theory.   This then provides
the basis for the statements of the quantum singularity conjecture
and the quantum cosmic censorship conjecture.

\subsection{Singularities in the classical observable algebra}

In section 3 we found an evolution equations for classical observables
in the gauge defined by (19) and (21).
We then used this to define both gauge
fixed and Dirac observables that correspond to making measurements
on the surface defined by $T(x)= \tau  $, given that the
clocks are synchronized by setting $T(x)=0$ on the initial
surface.   In this section I would like to discuss what effects
the  singularities of classical general relativity have on these
observables\footnote{As in section 4, I will in this section use
the notation
$phys$ to refer either to the gauge fixed or the Dirac formalisms.}.

Let us begin with a simple
point, which is the following:  {\it Given
that the matter fields, including the time fields, satisfy the positive
energy condition required by the singularity theorems, for
any $\tau_0$
there are  regions, ${\cal R}(\tau_0)$ of
the  phase space $\Gamma = \{ A,E, \phi ,...) $ such that
the future evolution of any data in this region becomes
singular before
the physical time $\tau_0$ (defined by the gauge conditions (19)
and (21)).  This means that the evolutions of the data in
${\cal R}(\tau_0)$
do not have complete $T(x)=\tau$ surfaces.}
This happens because, roughly speaking, some of the
clocks that define the $T(x)=$constant surfaces have
encountered spacetime singularities on surfaces with $T(x)<\tau_0$.

Now, let us consider what I will call "quasi-local"
observables ${\cal O}(\tau )_{phys}$, such as
${\cal A}^I(\tau )_{phys}$ and
$T^I(\tau )_{phys}$, which are associated with more or
less local regions of the initial data surface.
We may expect that for such observables
the following will be the case:   For each such ${\cal O}(\tau )_{phys}$
and for each $\tau_0 $ there will be  regions on
the phase space $\Gamma $ on which
 ${\cal O}(\tau_0 )_{phys}$ is not defined because, for data in that
region, the local region measured by that observable (picked
out by the spatial reference frame) encountered a curvature
singularity  at some time
$\tau_{sing}<\tau_0$.

Thus, it is clear that the existence of spacetime singularities does
limit
the operational notion of time tied to a field of clocks.  The limitation
is that, if we define the evolution of physical quantities in terms of
the physical observables
${\cal O} (\tau )_{phys}$,  only the $\tau=0$
observables that measure the properties of the initial data surface
can
be said to give good coordinates for the full space of solutions.   If we
want a complete description of the full space of solutions (defined as
the evolutions of non-singular initial data) we cannot get complete
information from the evolving constants of motion for any
nonzero $\tau$.  What we cannot get is complete
information about those solutions for which by
$\tau$ some of the clocks have already fallen into singularities.

As far as the classical theory is concerned, this limitation
is necessary and entirely unproblematic. The observables become ill
defined because we cannot ask any question about what is seen by
observers after they have ceased to exist.  As general relativity
is a local theory, if we choose our observables appropriately,
we can still have complete information about
all measurements that can be made at a time $\tau$ by local observers
who have not yet fallen into a singularity.

I now turn to a consideration of the implications of this for
the quantum theory.

\subsection{Singularities in quantum observables}

We have seen how the existence of singularities in classical theory
is expressed in terms of the classical observables ${\cal O}(\tau)_{phys}$.
There are now two  question that must be asked:   First,
in principle can singularities show up in
the same way in the physical operators of a quantum cosmological
theory?  Second, do they actually occur in the physical operator
algebra of
a realistic theory of quantum gravity coupled to matter fields?

 The
answer to the first question is yes, as has been shown in two model
quantum
cosmologies.  These are a finite dimensional example, the Bianchi I
quantum
cosmology\cite{ATU} \footnote{Similar
phenomena occur for other Bianchi models\cite{ATU}.}
and an infinite dimensional field theoretic example, the
one polarization Gowdy quantum cosmology\cite{viqar-time}.
These are both exactly
solvable systems; the first has a physical Hilbert space isomorphic to
the state space of a free relativistic
particle in $2+1$ dimensional
Minkowski spacetime while the Hilbert space of the second is
isomorphic
to that of a free scalar field theory in $1+1$ dimensions.  However,
in spite
of the existence of these isomorphisms to
manifestly non-singular
physical
systems, they are  each
singular theories when considered in terms of the
operators that represent observables of the corresponding
cosmological
models.  In both cases there is a global notion of time, which is the
volume of the universe, $V$  in a homogeneous slicing.
One can then
construct a
physical observable, called $C^2(V)$ which is defined as
\f
C^2(V) \equiv \int_{\Sigma (V)} \sqrt{q}
g^{\mu \nu} g^{\alpha \beta }
C^\rho _{\sigma \mu \alpha }C^\sigma _{\rho \nu \beta }
\ff
where $\Sigma (V)$ is the three surface defined in the slicing by the
condition that the spatial volume is
$V$ and $C^\sigma _{\rho \nu \beta }$ is the Weyl curvature.
There is also in each case, a
$V$-time dependent Hamiltonian that governs the evolution of
operators
such as $C^2(V)$ through Heisenberg equations of motion.

Now, it is well known that
in each of these models
the cosmological
singularity of the classical theory shows up in the fact that
$\lim_{V \rightarrow 0} C^2(V)$  is infinite.
What is, perhaps, surprising  is that the quantum
theory is equally singular, in that
$\lim_{V\rightarrow 0} \hat{C}^2(V)$
diverges.  The exact meaning of this is slightly different
in the finite dimensional and the quantum field theoretic
examples.  In the Bianchi I case, it has
been shown by
Ashtekar, Tate and Uggla\cite{ATU} that for any
two normalizable
states $|\Psi>$ and $|\Phi>$ in the physical Hilbert space
\f
\lim_{V\rightarrow 0} <\Psi |\hat{C}^2(V)|\Phi >_{physical} = \infty
\ff
In the one polarization Gowdy model,
Husain \cite{viqar-time} has shown that (given a physically
reasonable ordering for $\hat{C}^2(V)$) there is a unique
state $|0>$ such that (72) holds for all $|\Psi> $ and $|\Phi>$ which
are
not equal to $|0>$.  Furthermore,  for all $V$, $<0|\hat{C}^2 (V)|0>=0$
so the state $|0>$ represents the vacuum in which no degrees of freedom
of the gravitational degrees of freedom are ever excited.  As the
Gowdy
cosmology contains only gravitational radiation, this corresponds to
the
one point of the classical phase space which is just flat spacetime.
For any other states, the quantum cosmology is singular in the sense
that
the matrix elements of the Weyl curvature squared diverge at the
same
physical time that the classical singularities occur.

Thus, we see from these examples that there is no principle that
prevents
spacetime singularities from showing up in quantum cosmological
models and that they manifest themselves in the physical quantum
operator algebra in the same way they do in the classical observable
algebra.
It is therefore a dynamical question, rather than a question of
principle,
whether or not singularities can occur in the full quantum
theory.   While it is, of course, possible that
the singularities are eliminated in every consistent quantum
theory of gravity, I think it must be
admitted that at present there is little evidence that this is the case.
The evidence presently available about the elimination of
singularities
is the following: a)  Cosmological singularities are not eliminated in
the semiclassical approximation of quantum cosmology
\cite{semiclass}.
b)  There are exact solutions of string theory which are singular
\cite{gary-bh}.  c)  In $1+1$ models of quantum gravity, singularities
are sometimes, but not always,
eliminated, at the semiclassical
level\cite{witten-bh,gary-bh,andy-bh,preskill-bh}.

Furthermore, there does not seem to be any reason
why quantum cosmology requires the removal of the singularities.
Both the mathematical structure and the physical interpretation
of the quantum theory are, just like those
of the classical theory, robust enough to survive the occurrence of
singularities.

Given this situation, it is perhaps reasonable to ask
whether and how singularities may appear in full quantum gravity.
As a step towards answering this question, we may postulate the
following conjecture:

\blankline
\noindent{\bf Quantum singularity conjecture} {\it  There exist, in
the
Hilbert space ${\cal H}_{phys}$ of quantum gravity,
normalizable states $|\Psi >$
such that:
\blankline
\noindent
 a)  the expectation values
$<\Psi | \hat{\cal O}_I(0)_{phys} |\Psi >_{phys}$ are
finite for all $I$, so that
at the initial time $\tau=0$, all physical observables are finite.}
\blankline
\noindent
{\it b)  There is a subset of the physical
operators, $\hat{\cal O}_I(\tau )_{phys}$, whose
expectation values in the state $|\Psi >$ develops singularities
under evolution in the physical time $\tau$.  That is, for each
such $|\Psi>$ and  for each $\hat{\cal O}_I(\tau )_{phys}$ in this
subset
there is a finite time $\tau_{sing}$ such
that }
\f
\lim_{\tau \rightarrow \tau_{sing}} <\Psi |
\hat{\cal O}^I (\tau )_{phys}|\Psi >_{phys} = \infty
\ff

This means that if we want to predict what observers in the universe
described by the state $|\Psi>$ will
see at some time $\tau_0$, they may measure only the
$\hat{\cal O}_I(\tau_0)_{phys} $ which do not go singular in this
sense
by the time $\tau_0$.   This means that they may be able to recover
less
information about the state $|\Psi>$ by making measurements
at that time than was available
to observers at the time $\tau=0$.  Thus, the occurrence of
singularities
in the solutions to the operator evolution equations (43) (or (51))
means that
real loss of information happens in the full quantum theory.

The possibility of this happening can be captured by a conjecture
that I will call the {\it quantum cosmic censorship conjecture.}  The
name is motivated by analogy to the classical conjecture: if there is
censorship then there is missing information.  This is the
content of the following:

\blankline
\noindent
{\bf Quantum cosmic censorship conjecture:}   a)  {\it  There exists
states $|\Psi>$
in ${\cal H}_{phys}$ which are singular for at least one
observable
${\cal O}_{sing} (\tau )_{phys} $ at some time $\tau_{sing}$,
but for which
there are a countably
infinite number of other observables
${\cal O}^\prime_{I}(\tau)$ such
that $<\Psi | {\cal O}_{I} (\tau )_{phys} |\Psi > $ are well
defined and are finite for some
open interval of  times $\tau>\tau_{sing}$ .}
\blankline
\noindent
b) {\it Let $|\Psi>$ be a
state which satisfies these conditions.  Then for  every $\tau$
in this open interval
there is a proper density matrix
$\rho_{\Psi}(\tau) $ such that,  for every  $ \hat{\cal O}_I (\tau) $,
for which  \newline
$<\Psi | \hat{\cal O}_I (\tau) |\Psi > $ is finite, then}
\f
<\Psi | \hat{\cal O}_I (\tau) |\Psi >_{phys} = Tr
\rho_{\Psi }(\tau )  \hat{\cal O}_I (\tau) .
\ff
Here the trace is to be defined with respect to the physical inner
product a proper density matrix is one that corresponds to no
pure state.

The first part of the conjecture means that there are states which
describe
what we might want to call
black holes in the sense that while some observables become
singular
at some time $\tau_{sing}$ there are other observables which remain
nonsingular for later times. The second part means that
there exists a  density matrix that
contains all the information about
the quantum state that is relevant for physical
times $\tau$ after the
time of the first occurrence of singularity.
Because a density matrix contains all the
information that could be gotten by measuring the pure state, we
may say that loss of information has occurred.

Finally, we may note that for any state $|\Psi >$ there may
be a finite time $\tau_{final}$ such that, for every observable
${\cal O}_I (\tau )_{phys}$,
$ <\Psi | {\cal O}_I (\tau)_{phys} |\Psi >_{phys} $ is undefined, divergent
or zero for
every $\tau> \tau_{final}$.  This would correspond to a quantum
mechanical
version of a final singularity.

Suppose these conjectures can be proven for quantum general
relativity,
or some other quantum theory of gravity.  Would this mean that the
theory would be inadequate for a description of nature?  While
someone may want to argue that it may be preferable to have a
quantum
theory of gravity without singularities, I do not think an argument
can
be made that such a theory must be either incomplete, inconsistent
or in
disagreement with anything we know about nature.  What
self-consistency and consistency with observation require
of  a quantum theory of cosmology is much less.  The following may
be
taken to be a statement of the minimum that we may require of a
quantum
theory of cosmology:

\blankline
\noindent
{\bf Postulate of adequacy}.  A quantum theory of cosmology,
constructed
within the framework described in this paper, may be called {\it
adequate}
if,
\blankline
\noindent
a)  {\it For every $\tau>0$ there exists a physical state
$|\Psi > \in {\cal H}_{phys}$ and a countable set of operators
$\hat{\cal O}_I (\tau)$ such that the
$<\Psi |\hat{\cal O}_I (\tau)_{phys}|\Psi>_{phys} $ are finite.}

\blankline
\noindent
b)  {\it  The theory has a flat limit, which is quantum field
theory on Minkowski spacetime.  This means that there
exists physical states and operators whose expectation
values are equal to those  of quantum field theory on Minkowski
spacetime for large  regions of space
and time, up to errors which are small in Planck units.}

\blankline
\noindent
c)  {\it The theory has a classical limit, which is
general relativity coupled
to some matter fields.  This means that there exists physical
states and operators
whose physical expectation values are equal to the values of the
corresponding classical observables evaluated in a classical solution
to general relativity, up to terms that are small in Planck units.}

If a theory satisfies these conditions, we would have a great deal of
trouble saying it was not a satisfactory quantum theory of gravity.
Thus, just like in the classical theory, the presence of singularities
and
loss of information cannot in principle prevent a quantum theory of
cosmology from providing a meaningful and adequate description of
nature.
Whether there is an adequate quantum theory of cosmology that
eliminates
singularities and preserves information is a dynamical
question, and whether that theory,  rather than
another adequate theory for which the quantum singularity and
quantum cosmic censorship conjectures hold, is the correct
description
of nature is, in the end, an empirical question.

\section{Conclusions}

The purpose of this paper has been to explore the implications
of taking a completely pragmatic approach to the problem of
time in quantum cosmology.  The main conclusion of the
developments described here is that such an approach may be
possible at the nonperturbative level.  This may allow the theory
to address problems such as the effect of quantum effects on
singularities which most likely require a nonperturbative
treatment, while remaining within the framework of a coherent
interpretation.

I would like to close this paper by discussing two questions.  First,
are there ways in which the ideas described here may be tested?
Second, is it possible that there is a more fundamental solution to the
problem of time in quantum gravity which
avoids the obvious limitations
of this pragmatic approach?

\subsection{Suggestions for future work}

The proposals and conjectures described here are only meaningful to
the
extent that they can be realized in the context of a full
quantization
of general relativity or some other quantum theory of gravity.
In order to
do this, the key technical problem that must be resolved is, as we
saw in
section 3, the construction of the operator ${\cal W}$.  Given that we know
rather a lot about both the kernel and the
action of the Hamiltonian constraint, I believe that this is
a solvable problem.

Beyond this, it would be very interesting to test these ideas and
conjectures
in the context of certain model systems.  Among those that could be
interesting are 1)  The Bianchi IX model  2)  The full two polarization
Gowdy models 3)  Models of spherically symmetric general relativity
coupled to matter\footnote{For the complete quantum theory
without matter, see \cite{thomas}.} 4)  Other $1+1$ dimensional
models of quantum
gravity coupled to matter such as the dilaton theories that have
recently received some
attention \cite{CGHS,gary-bh,andy-bh,banks-bh,lenny-bh,hawking-2d}.
5)  the chiral $G\rightarrow 0$
limit of the theory\cite{Gtozero} and finally, 6)
$2+1$ general relativity coupled to
matter\cite{witten-2+1,2+1-us}.  Each of these are systems
that have not yet been solved,
and in which the difficulty of finding the physical observables has,
as in the full theory, blocked progress.

\subsection{Is there an alternative framework for quantum
cosmology
not based on such an operational notion of time?}

We are now, if the above is correct, faced with the following
situation.
Taking an operational approach to the meaning of time
 we have been able to provide a complete physical
interpretation for quantum cosmology that reduces to quantum
field theory in a suitable limit.  However,
the quantum field theory that is reproduced may turn out
to be  unitary only
in the approximation in which we can neglect the possibility that
some of the clocks that define operationally surfaces of simultaneity
become engulfed in black hole singularities.  This need not be
disturbing;
it says that we cannot count on probability conservation in time if the
notion of time we are thinking of is based on the existence of a
certain
field of dynamical clocks and there is finite probability that these
clocks themselves cease to exist.
However, if there is no other notion of time with respect to
which probability conservation can be maintained, so that
this is the best that can be done,  it is still a bit disturbing.

We seem at this juncture to have two  choices.  It
may indeed
be that we cannot do better than this, so that we must accept that
the Hilbert space structure that forms the basis of our interpretation
of
quantum cosmology is tied to the existence of
certain physical frames of
reference and that, as the existence of the conditions that define
these
frames of reference is contingent, we cannot ascribe
any further meaning to unitarity.  If this is the case then we have to
accept a further "relativization" of the laws of physics, in which
different
Hilbert structures, with different inner products, are associated to
observations made by different observers.  This
means, roughly, that not only are the actualities (to use a distinction
advocated by Shimony\cite{abner}) in quantum mechanics
dependent
on the physical conditions of the observer, so are the potentialities.
This point of view
has been advocated by Finkelstein\cite{david} and developed
mathematically
in a very interesting recent paper of Crane\cite{louis-time}.

On the other hand, it is possible to imagine that there is some
meaning
to what the possibilities are for actualization that is independent of
the
conditions of the observer.  If such a level of the theory
existed, it could
be used to deduce the relationships between what could be, and
what
is, seen by different observers in the same universe.

Barbour has recently made a proposal about the role of time in
quantum cosmology, which I think
can be understood along these lines\cite{julian-heap}.
I would now like to sketch it,  as it may serve as a prototype for all
such proposals in which the probability interpretation of quantum
cosmology is not relativized so as to make the inner produce
dependent
on the conditions of the observer.

Barbour's proposal is at once a new point of view about time and a
new
proposal for an interpretation of quantum cosmology.  He posits that
time actually does not exist, so that our impressions of the existence
and passage of time are illusions caused by certain properties of the
classical limit of quantum cosmology.  More precisely, he proposes
that an interpretation can be given entirely in the context of
the diffeomorphism
invariant theory.

Barbour's fundamental postulate, to which he gives the colorful
name of
the "heap hypothesis", is as follows: {\it The world consists
of a timeless
real ensemble of configurations, called "the heap".}  The probability
for any given spatially diffeomorphism invariant
property to occur in "the heap" is governed by a quantum state,
$|\Psi >$ which
is assumed to satisfy all the constraints of quantum gravity.  This
 probability is considered to be
an actual ensemble average.   Given a particular diffeomorphism
invariant observable $\hat{\cal O}_I$, the ensemble expectation
value in the heap is given by
\f
<\Psi | \hat{\cal O}_I|\Psi >_{complete \ diffeo}   .
\ff
Here the inner product is required to be the spatially
diffeomorphism
invariant inner product for the whole system, including any clocks
that may be around.  Thus, this proposal is different than the one
made in section 3 in which the inner product proposed in (46)
is the spatially diffeomorphism invariant inner product
for a specifically reduced system in which the clock has been
removed.

There are several comments that must be made about this
proposal.  First, this is the complete statement of the interpretation
of the theory.  Quantum cosmology
is understood as giving a statistical description
of a real ensemble of configurations, or moments.  There is no
time.  The fact that we have an impression of time's passage is entirely
to be explained by certain properties of the quantum state of the
universe.  In particular, Barbour wants to claim that our experience
of each complete moment is, so to speak, a world unto itself.  It is
only because we have memories that we have an impression in this
moment that there have been previous moments.  It is only because
the quantum state of the universe is close to a semiclassical state in
which the laws of classical physics approximately hold that the world
we experience at this moment gives us the strong impression of
causal
connections to the other moments.

Second, the probabilities given by (75)
are not quantities that are necessarily or directly
accessible to observers like ourselves who live inside the
universe.  Only
an observer who is somehow able to look at the whole ensemble
is able to
directly measure the probabilities given by (75).
Of course, we are not in that situation.  Barbour must then explain
how the probabilities for observations that we make are related to
the
 probability distribution for elements of the heap to have
 different properties.  In order to do this, the key thing that he must
 do is show how
the probabilities defined by the heap ensemble (75) are related
to what we measure.

One way in which this may happen is that if one considers only
states
in which some variables corresponding to a
particular clock are semiclassical,
then Barbour's inner product (75) may reduce to the inner product
defined by (46) in which the clock degrees of freedom have
been removed.  From Barbour's point of view, the inner product
proposed in this paper could only be an approximation to the true
ensemble probabilities (75) that holds in the case that the quantum
state of the universe is semiclassical in the degrees of freedom of
a particular physical clock.

This discussion of Barbour's proposal brings us back
to the choice I mentioned above and
points up what I think is a
paradox that must confront any quantum theory of cosmology.
The two possibilities we must, it seems, choose between can
be described as follows:  a)  The
inner product and the resulting probabilities are tied in an
operational
sense to what may be seen by a physical observer inside the
universe.  In
this case, as we have seen here, the notions of unitarity and
conservation
of probability can only be as good as the clocks carried by a
particular
observer may be reliable.  We are then in danger of the kind of
relativization of the interpretation mentioned above.  b)  The
basic statement of the interpretation
refers neither to a particular set of observers nor to time as
measured by
clocks that they carry.  In this case the relativization can be
avoided, but
at the cost that the fundamental quantities of the theory do not
in general refer to
any observations made by observers living in the universe.  In this
case the probabilities seen by any observers living in the universe
can only approximate the true probabilities for particular
semiclassical configurations.  Furthermore, if there is a finite
probability that any physical clock may in its future encounter a
spacetime singularity, however small, then it is difficult to see
how a breakdown of unitarity evolution can be avoided, if
by evolution we mean anything tied to the readings of a physical
clock.

This situation brings us back to the problem of what happens to
the information inside of an evaporating
black hole.   I think that the minimum that can be deduced
from the considerations of section 6 is that, at the nonperturbative
level, this question cannot be resolved without resolving the
dynamical
question of what happens to the singularities inside classical
black holes.   As pointed out a long time ago by Wheeler\cite{MTW},
this problem challenges all of our ideas about short distance physics
and its relation to cosmology\footnote{One speculative proposal
about this is in \cite{evolve}.}.  Unfortunately, it must be admitted
that the quantum theory of gravity still has little to say about this
problem.      What I hope to have shown here is that there may be
a language which allows the problem to be
addressed by a nonperturbative formulation of the
quantum theory.  Whether it can be answered by such a formulation
remains a problem for the future.

\section*{ACKNOWLEDGEMENTS}

This work had its origins in my attempts over
the last several years to understand and resolve issues
that arose in collaborations and discussions with Abhay Ashtekar,
Julian
Barbour, Louis Crane, Ted Jacobson and Carlo Rovelli.
I am grateful to them
for continual stimulation, criticism and company on this long road
to quantum gravity.  I am in addition indebted to Carlo Rovelli
for pointing out an error in a previous version of this
paper.  I am also very grateful to a number of other
people
who have provided important stimulus or criticisms of these ideas,
including
Berndt Bruegmann, John Dell,  David Finkelstein, James Hartle,
Chris Isham, Alejandra Kandus, Karel Kuchar, Don Marolf, Roger
Penrose,
Jorge Pullin, Rafael
Sorkin, Rajneet Tate and John Wheeler.
This work was supported by the National
Science Foundation under grants PHY90-16733 and
INT88-15209 and
by research funds provided by Syracuse University.

\bibliographystyle{plain}

\end{document}